\documentclass[iop]{emulateapj}
\usepackage{epstopdf}
\usepackage{longtable}
\usepackage{mathtools}

\begin{document}

\title{The GALEX Nearby Young-Star Survey}

\author{David R. Rodriguez\altaffilmark{1}, B.\ Zuckerman\altaffilmark{2}, Joel H.\ Kastner\altaffilmark{3}, M.\ S.\ Bessell\altaffilmark{4}, Jacqueline K.\ Faherty\altaffilmark{1,5}, Simon J.\ Murphy\altaffilmark{4,6} }

\altaffiltext{1}{Departamento de Astronom\'ia, Universidad de Chile, Casilla 36-D, Santiago, Chile \\
(drodrigu@das.uchile.cl)}
\altaffiltext{2}{Dept.\ of Physics \& Astronomy, University of California, Los Angeles 90095, USA}
\altaffiltext{3}{Center for Imaging Science, Rochester Institute of Technology, 54 Lomb Memorial Drive, Rochester NY 14623}
\altaffiltext{4}{The Australian National University, Cotter Road, Weston Creek ACT 2611, Australia}
\altaffiltext{5}{Department of Astrophysics, American Museum of Natural History, Central Park West at 79th Street, New York, NY 10034}
\altaffiltext{6}{Gliese Fellow, Astronomisches Rechen-Institut, Zentrum f\"{u}r Astronomie der Universit\"{a}t Heidelberg, Germany 69120}

\begin{abstract}

We describe a method that exploits data from the GALEX ultraviolet and WISE and 2MASS infrared source catalogs, combined with proper motions and empirical pre-main sequence isochrones, to identify candidate nearby, young, low-mass stars. Applying our method across the full GALEX-covered sky, we identify 2031 mostly M-type stars that, for an assumed age of 10 (100) Myr, all lie within $\sim$150 ($\sim$90) pc of Earth. The distribution of M spectral subclasses among these $\sim$2000 candidate young stars peaks sharply in the range M3--M4; these subtypes constitute 50\% of the sample, consistent with studies of the M star population in the immediate solar neighborhood. 
We focus on a subset of 58 of these candidate young M stars in the vicinity of the Tucana-Horologium Association. Only 20 of these 58 candidates were detected in the ROSAT All-Sky X-ray Survey --- reflecting the greater sensitivity of GALEX for purposes of identifying active nearby, young stars, particularly for stars of type M4 and later. Based on statistical analysis of the kinematics and/or spectroscopic followup of these 58 M stars, we find that 50\% (29 stars) indeed have properties consistent with Tuc-Hor membership, while 12 are potential new members of the Columba Association, and two may be AB Dor moving group members. 
Hence, $\sim$75\% of our initial subsample of 58 candidates are likely members of young (age $\sim$10--40 Myr) stellar moving groups within 100~pc, verifying that the stellar color- and kinematics-based selection algorithms described here can be used to efficiently isolate nearby, young, low-mass objects from among the field star population. 
Future studies will focus on characterizing additional subsamples selected from among this list of candidate nearby, young M stars.

\end{abstract} 


\section{Introduction}

The past two decades have seen the discovery of a number of young (age$<$100~Myr) 
stellar moving groups and associations within $\sim$100~pc of Earth (for reviews see 
\citealt{ZS04} and \citealt{Torres:2008}). By studying stars in these moving groups one 
can observe the evolution of stellar properties as a function of juvenile ages. Nearby 
young stars and brown dwarfs are excellent targets for 
studies of gas-rich circumstellar disks (e.g., \citealt{Qi:2004,Qi:2006,Qi:2008,Hughes:2008,Kastner:2008, 
Rodriguez:2010,Zuckerman:2012}) and direct imaging surveys of young Jupiters (e.g., \citealt{Chauvin:2004,Chauvin:2005,Marois:2008,Marois:2010,Lagrange:2010}). 
At present, the initial mass functions of these young stellar groups are 
poorly determined for spectral types later 
than $\sim$M2 (e.g., \citealt{Torres:2008,Shkolnik:2011,Schlieder:2012a}). 
In contrast, the field population of M dwarfs peaks strongly at spectral types M3--M5 
(see \citealt{Farihi:2005,Henry:2006,Reid:2007,Bochanski:2010,Stelzer:2013}, and references therein) suggesting a 
significant number of nearby, young stars with spectral types later than M3 remains to 
be discovered. Identification of additional young moving group members, particularly low-mass 
stars, is a key step both in our understanding of the processes involved in the early 
stages of stellar and substellar evolution and in searches for recently formed planets.

Given the incompleteness of the young moving groups at the low-mass end, many programs are 
actively searching for these ``missing" M-dwarfs with a variety of techniques (e.g., see 
\citealt{Rodriguez:2011,Riedel:2011,Shkolnik:2011,Shkolnik:2012,Schlieder:2012a,Malo:2012}; 
and references therein). Several techniques exploit the ultraviolet and X-ray properties 
of low-mass stars. As low-mass (M$<$1~$M_\odot$) pre-main sequence stars descend to the 
main sequence, their deep convective envelopes combine with differential rotation to 
produce strong magnetic dynamos. This stellar dynamo generates high levels 
of chromospheric and coronal activity; the former is a source of ultraviolet (UV) 
emission while the latter contributes to X-ray emission 
\citep{Linsky:2001,Preibisch:2005,Stelzer:2013}. Recent work \citep{Findeisen:2011, 
Rodriguez:2011, Shkolnik:2011, Findeisen:2010} has linked stellar activity due to youth 
with UV emission detected with the Galaxy Evolution Explorer satellite (GALEX; 
\citealt{Martin:2005}). 
\citet{Shkolnik:2011} suggest that such UV-bright low-mass stars have ages $\sim$300~Myr or 
younger, based on the strength of their X-ray emission (see Section~6 in \citealt{Shkolnik:2011}).
These low-mass stars are brightest at near-infrared wavelengths 
and thus are best identified with surveys such as the Two-Micron All Sky Survey (2MASS; 
\citealt{Cutri:2003}). The recent release of the Wide-field Infrared Survey Explorer 
catalog (WISE; \citealt{Wright:2010}) has opened up the sky at additional near- to 
mid-infrared wavelengths beyond those covered with 2MASS, facilitating the 
identification of cooler and fainter low-mass objects. The bright UV emission from the 
stellar chromosphere, combined with cool-star colors at optical and infrared 
wavelengths, make the GALEX source catalog a powerful tool to search for nearby, young, 
low-mass stars.

Our prior work tying together UV emission with near-IR photometry was carried out for 
the Scorpius-Centaurus Region and TW~Hya Association (TWA; \citealt{Rodriguez:2011}). 
This study introduced a method to identify candidate young stars via the combination of their (a) 
UV-IR colors, (b) estimated range of motion in the context of Galactic space velocities for 
known young stellar groups, and (c) spectral signatures 
of youth. In the present work, we extend our methodology to incorporate the WISE data, 
which permits characterization of the full spectral energy distribution of UV-bright stars, 
allows estimation of proper motions (given the $\sim$10 year baseline between WISE 
and 2MASS), and breaks the JHK color degeneracy for early- to mid-M dwarfs. 
We have also now extended our initial study to the entire GALEX-covered sky (61\% of the total sky), allowing detection of thousands of candidate young low-mass stars, some of which may be members of previously undiscovered moving groups. 
We have also devised an algorithm to identify such potential stellar groups 
based on their similar proper motions and estimated distances (Section~\ref{fvalue}) 
and have carried out a convergent point analysis for several young 
moving groups (Section~\ref{cpanalysis}). 
We refer to these collected efforts to identify young stars through the combination of GALEX and other catalog data as the GALEX Nearby Young-Star Survey (GALNYSS).
Here, we describe our methodology and present the first GALNYSS results, with a focus on the Tucana-Horologium Association.

\section{GALNYSS Methodology}\label{method}

We have developed a search strategy that consists of combining the GALEX, 2MASS, and WISE catalog 
information to identify nearby, young, low-mass stars. This is similar to the work presented in 
\citet{Rodriguez:2011}, but revised to utilize WISE data as well as improved isochrones and selection 
criteria. The steps in our identification sequence are as follows:

\begin{enumerate}
\item Combine IR and UV photometry to identify objects likely to be nearby, M-type dwarfs with UV 
excesses (Section~\ref{method:uv}).
\item Use cataloged and calculated proper motions (Section~\ref{pminfo}), combined with estimates of 
spectral types and photometric distances (Section~\ref{spectypes}), to select the subset of these objects 
that are candidate M-type stars within $\sim$150~pc (if 10~Myr old) and have UVW space motions similar to those of known 
young stellar groups (Section~\ref{uvw}).
\item Perform spectroscopic followup of the best resulting candidates (Section~\ref{candidates}) to 
measure radial velocities along with spectral features, such as H$\alpha$ emission and Li absorption 
lines, that are diagnostic of youth (Section~\ref{specfollowup}).

\end{enumerate}

\subsection{GALEX (UV-based) Identification of Candidate Nearby, Young, Low-Mass Stars} \label{method:uv}

\begin{figure}[htb]
\begin{center}
\includegraphics[width=14cm,angle=0]{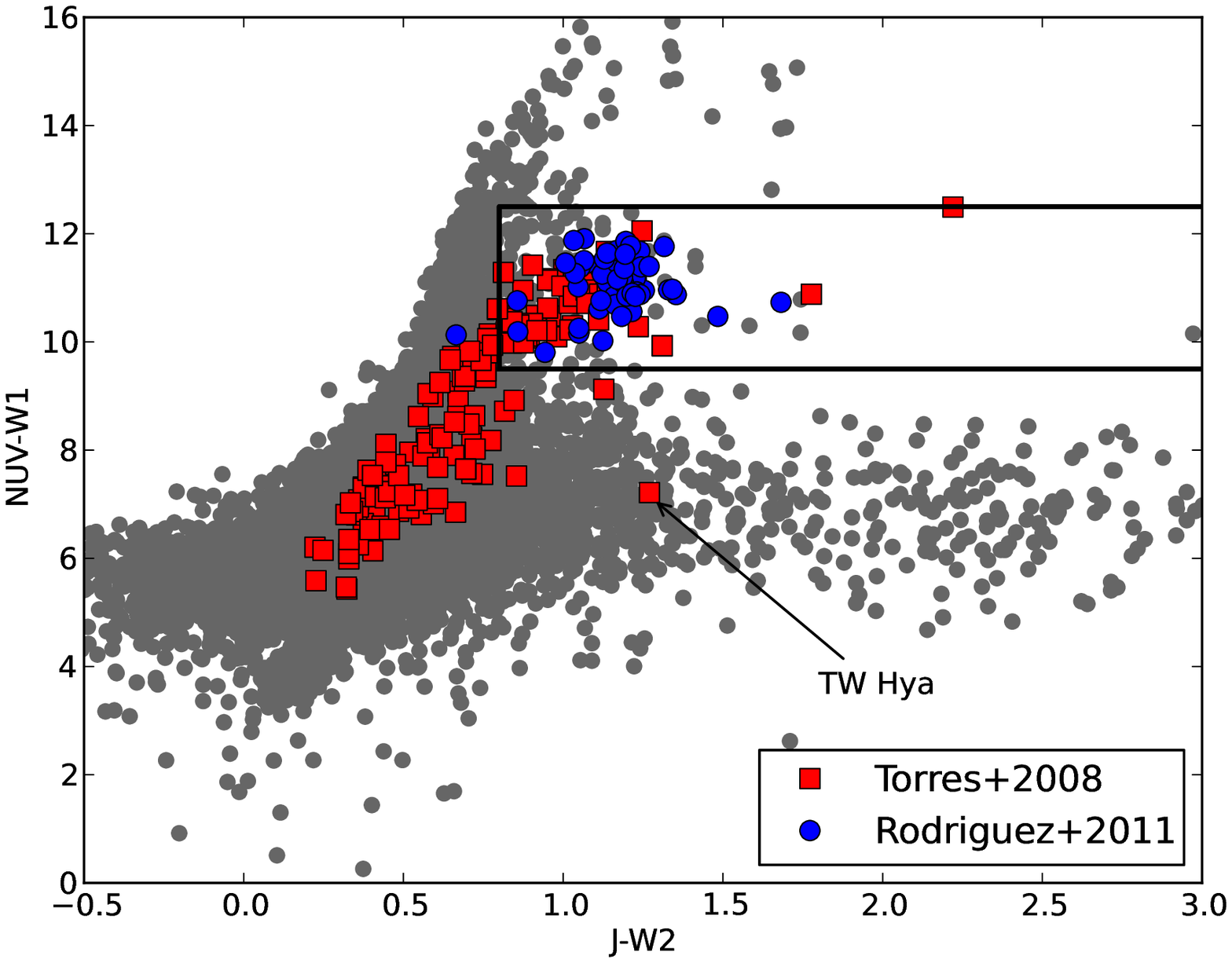}
\end{center}
\caption{NUV--W1 vs.\ J--W2 magnitudes for field sources detected in GALEX and WISE/2MASS as well as 
members (or candidates) of nearby, young moving groups \citep{Torres:2008,Rodriguez:2011}. 
The ordinate is NUV--W1 color from GALEX and WISE (W1 corresponds to 3.4 $\mu$m emission) and the abscissa is J--W2 color from 2MASS and WISE (W2 corresponds to 4.6 $\mu$m emission). 
The rectangle shows our color selection criteria (see Table~\ref{select_criteria} and Section~\ref{method:uv}).
Young stars with J--W2$>$1 are readily distinguished from older main sequence stars which have J--W2$<$1 and NUV--W1 between 10 and 14; J--W2 can be used a proxy for spectral type (see Equation 2 and Figure\ref{bd_jw2}). Distant UV-bright galaxies are the diffuse cloud of objects with NUV--W1 between 5--9 and J--W2$>$1. TW~Hya is the object located at J--W2$\sim$1.3 and NUV--W1$\sim$7; \citet{Rodriguez:2011} suggest its unique location in UV-IR colors is due to the 
accretion within the system. }
\label{jw2}
\end{figure}

To identify candidate young, low-mass stars we first examine where in 
various color-color combinations known young stars (as drawn from 
\citealt{Rodriguez:2011} and \citealt{Torres:2008}) lie with respect to other field GALEX sources. Figure~\ref{jw2} demonstrates this in a combined GALEX-WISE-2MASS 
NUV--W1 vs.\ J--W2 color-color diagram.  Here NUV is the near UV GALEX channel, W1 and W2 are the 3.4 and 
4.6 $\mu$m WISE bands and J is from the 2MASS catalog.  
As we demonstrate in Section~\ref{spectypes}, J--W2 color can be used as a proxy for spectral type. 
In Figure~\ref{jw2}, the young low-mass stars from \citet{Torres:2008} and \citet{Rodriguez:2011} stand out with respect to earlier-type young stars, older field stars, and the galaxy population. 
The J--W2 colors of young, low-mass stars are red and, while their NUV--W1 colors are bluer than those of field stars, they are not as blue as typical galaxies (NUV--W1 colors between 5 and 9).

Figure~\ref{jw2} drives our main selection criteria, which are listed in Table~\ref{select_criteria}. 
The first two criteria define where known young, low-mass stars are located in color-color space. 
Criterion \#3 is used to ensure our empirical spectral type relationship and isochrones are valid (see Section~\ref{spectypes}). 
Criterion \#4 (from the WISE catalog) ensures there is only 1 2MASS source associated with a given WISE source within 3\arcsec. This avoids 
misassociation of IR sources in crowded fields, though such confusion is worst near the Galactic plane, 
where GALEX has not observed. The remaining criteria are used to avoid saturated sources and to exclude 
contaminating, background galaxies. We note that members of young moving groups listed in 
\citet{Torres:2008} are removed from the GALNYSS candidate tables prior to our analysis, to avoid re-identifying 
known young stars.  
However, other lists of proposed young stars (e.g., \citealt{ZS04}, \citealt{Rodriguez:2011}) are not excluded.

\subsection{WISE-2MASS Proper Motions}\label{pminfo}

\begin{figure}[htb]
\begin{center}
\includegraphics[width=14cm,angle=0]{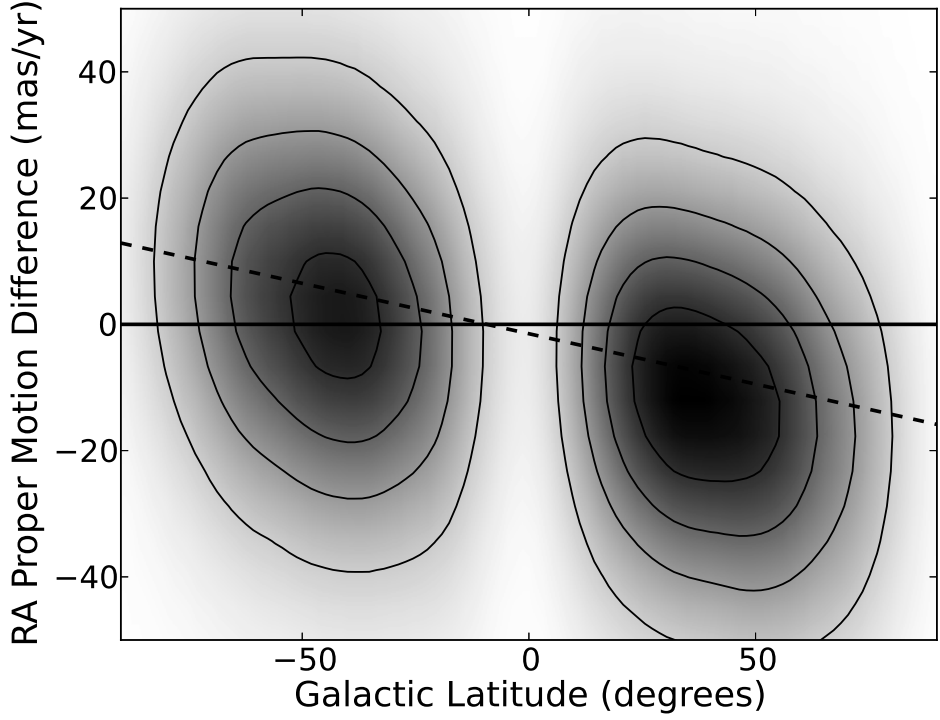}
\end{center}
\caption{Density plot of RA proper motion differences for 300,000 sources. WISE-2MASS proper motions are compared to PPMXL as a function of galactic latitude. A similar trend is observed when comparing to UCAC4 or USNO-B1. 
As these were originally drawn from our GALEX tables, no coverage exists for galactic latitudes $<$10 degrees. 
While a similar discrepancy can also be seen in Ecliptic longitude, it is corrected most easily in Galactic coordinates.
The dashed line indicates the correction used (Section~\ref{pminfo}).}
\label{fig:pm2}
\end{figure}

We make use of proper motion (PM) information to distinguish between candidate young stars and other UV excess sources, such as background galaxies or nearby flare stars. 
We have cross-correlated our sample against the UCAC4 \citep{Zacharias:2012} and PPMXL \citep{Roeser:2010} proper motion catalogs. In total, $\sim$56\% and $\sim$98\% percent of our color-selected candidates have matches with objects listed in these respective catalogs. 
Comparison with other catalogs shows that 92\% of the UV excess sources have proper motions in SuperCosmos \citep{Hambly:2001} and 70\% in USNO-B1 \citep{Monet:2003}.
For any source that lacked a published proper motion, a PM was estimated directly from the WISE-2MASS astrometry. With the $\sim$10-year baseline and 3\arcsec\, cross-correlation radius, these catalogs provide reasonable estimates of proper motion (up to $\sim$300~mas/yr) suitable for the first steps in our analysis. 
We note, however, a small ($\sim$10--15~mas/yr) systematic offset in the WISE-2MASS proper motions 
in the RA direction as a function of galactic latitude. Figure~\ref{fig:pm2} shows this offset when examining over 300,000 sources in the PPMXL catalog. Similar results are found when comparing with UCAC4 and USNO-B1. 
The WISE Explanatory Supplement\footnote{See Section 6.4 in\\ \url{http://wise2.ipac.caltech.edu/docs/release/allsky/expsup}} mentions that, given that the WISE astrometry is tied to that of 2MASS and no proper motion information was considered for the 10-year baseline, the WISE coordinate frame has a distortion relative to the International Celestial Reference System. A full correction is beyond the scope of this paper, but we attempt to remedy the RA offset with a linear term:
\begin{eqnarray}
\mu_{RA} = \mu_{RA_0} - 0.15963 \times b - 1.48284
\end{eqnarray}
Here, $\mu_{RA_0}$ and $\mu_{RA}$ are the proper motions in the RA direction before and after the correction.
We note that this offset is smaller than the uncertainty we describe below.

\begin{figure}[htb]
\begin{center}
\includegraphics[width=8cm,angle=0]{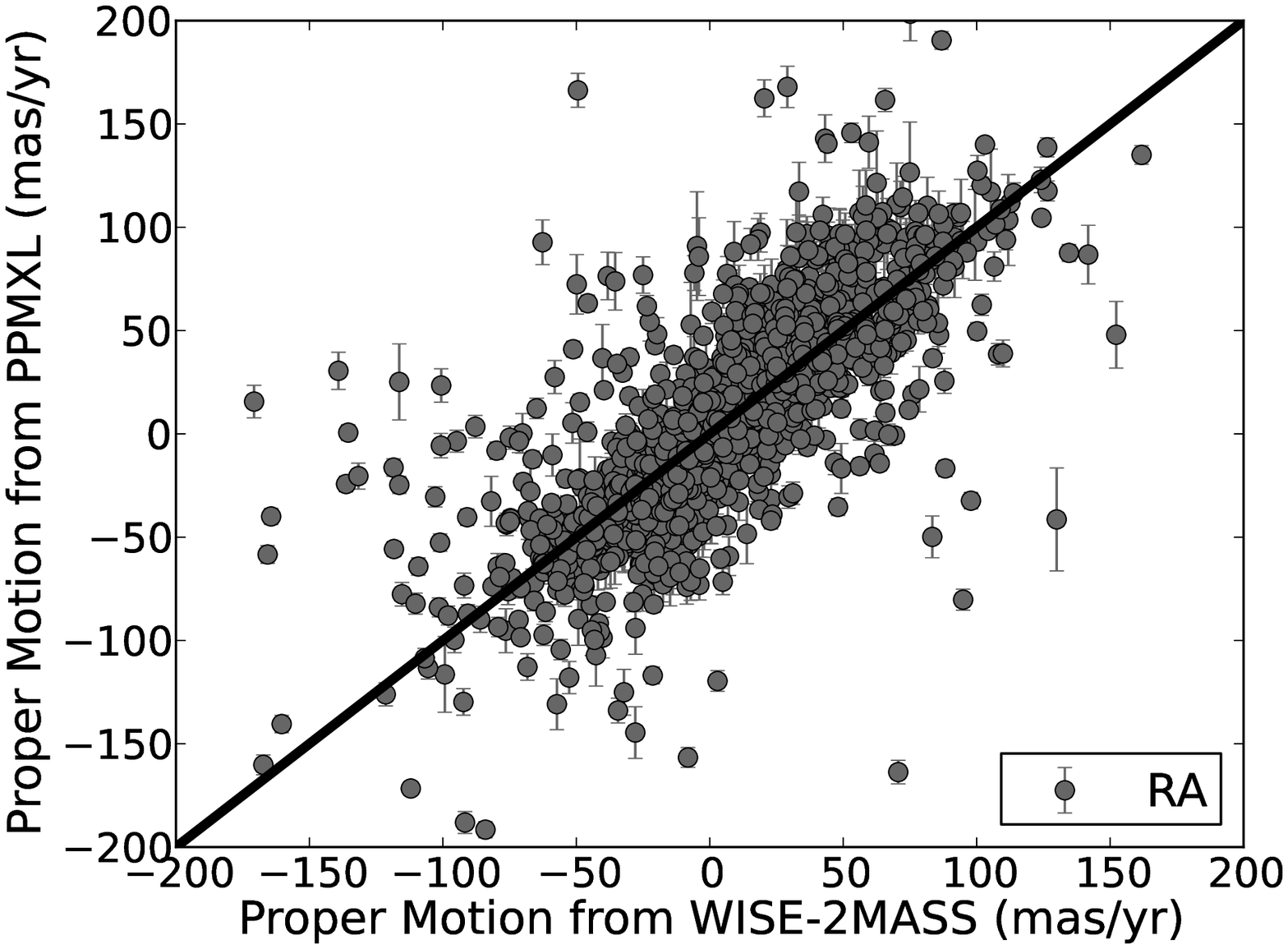}
\includegraphics[width=8cm,angle=0]{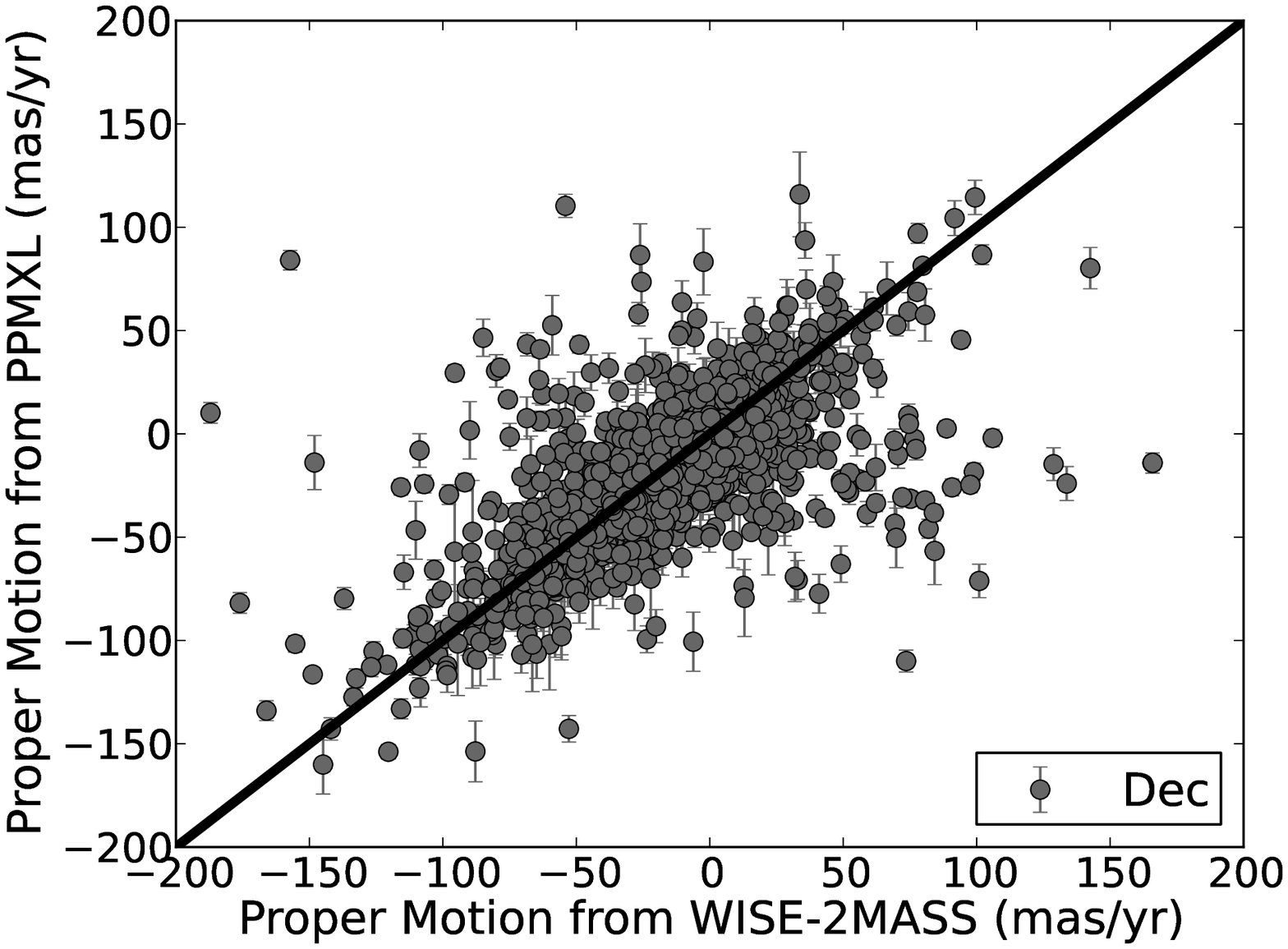}
\end{center}
\caption{Proper motions from PPMXL compared to those we have derived from WISE-2MASS astrometry for the stars in this study.  
The rms scatter is $\sim$25~mas/yr. 
Similar results are found when comparing WISE-2MASS proper motions against those listed in the UCAC3, USNO-B1, SuperComos, or UCAC4 catalogs. A small ($\sim$10--15~mas/yr) systematic offset in RA proper motion has been corrected (see Section~\ref{pminfo} and Figure~\ref{fig:pm2}).
}
\label{fig:pm}
\end{figure}

Figure~\ref{fig:pm} compares proper motions estimated from 
WISE-2MASS astrometry to those from PPMXL for the final list of $\sim$2000 candidates in this study (see Section~\ref{candidates}). 
From the rms scatter, we conservatively adopt 25~mas/yr as an estimate of the uncertainty for objects without proper motions in published catalogs.
We thus have at least one proper motion estimate for each candidate object, though in practice most objects have several estimates in good agreement with each other.

\subsection{Spectral Types and Distance Estimates} \label{spectypes}

\begin{figure}[htb]
\begin{center}
\includegraphics[width=14cm,angle=0]{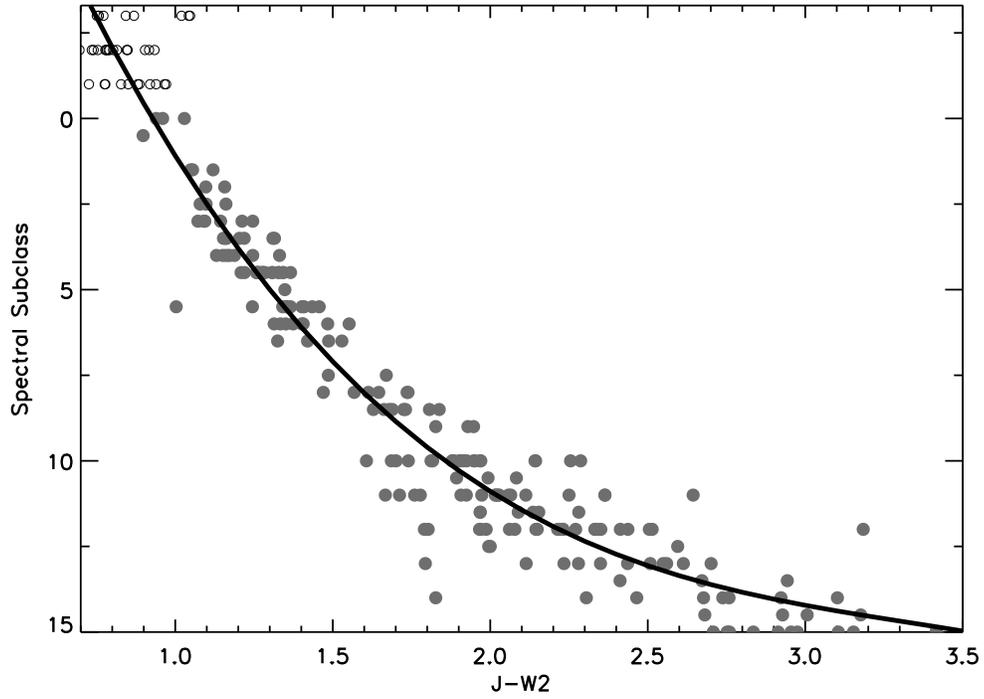}
\end{center}
\caption{Empirical spectral type relationship derived using WISE-2MASS photometry and MLT-dwarfs from 
\citet{Kirkpatrick:2011}. On the ordinate, 0=M0, 5=M5, 10=L0, and so forth. We include K-dwarfs (open circles; --1=K7, --2=K5, --3=K4) from \citet{Stauffer:2010}. 
We expect our GALEX sources to be mainly M dwarfs and anticipate finding few or no bonafide UV-emitting L dwarfs (see Section~\ref{spectypes}). The solid line is Equation (2) and applies best for late K to late M spectral types; see Section~\ref{spectypes} for details. }
\label{bd_jw2}
\end{figure}

Recently, \citet{Kirkpatrick:2011} presented a list of M, L, and T-dwarfs identified with WISE.
We have used their sample and the WISE-2MASS photometry to derive an empirical relationship tying $J-W2$ color to spectral type:
\begin{align*}
%\begin{eqnarray*}
&SpNum = -20.039 + 28.402 \times (J-W2)   &\text{\hfill (2)} \\
&- 8.084 \times (J-W2)^2 + 0.808 \times (J-W2)^3 
%\end{eqnarray*}
\end{align*}
SpNum is the number past M0, so 0=M0, 5=M5, 10=L0, and so forth (valid from K5/7--L5).
Figure~\ref{bd_jw2} shows the \citet{Kirkpatrick:2011} sources and our empirical relationship, which works best at spectral types late-K to late-M.
We note that this relationship is not valid for L types later than $\sim$L5 and thus impose a color limit of W1--W2$<$0.6, which selects objects earlier than L5.
Despite this limit, we do not anticipate identifying a substantial number of bonafide UV-bright L-dwarfs:
we have examined a list of 51 L-dwarf candidates within 40~pc (from \citealt{Kirkpatrick:2008}) with GALEX and find that, although 41 of them lie in GALEX-covered regions, none are detected. Similar results are found when searching among L and T dwarfs listed at DwarfArchives.org.

Candidates identified in the present program have spectral types predominantly in the M3--M5 range, as estimated from the relationship above.
These spectral types are, in general, accurate to $\pm$1 subclass in the early- to mid-M dwarf range, as we demonstrate in Section~\ref{specoverview}.
While some studies have presented redder colors for young L-dwarfs compared to older field dwarfs (see \citealt{Faherty:2013}, and references therein), this may not be the case for the M-dwarf spectral types we are sensitive to or to the J--W2 colors we use (see, for example, \citealt{Lyo:2004}).
Warm circumstellar material in the system can produce more emission in the WISE bands and thus make an object redder in J--W2 color, mimicking a later spectral type.
A visual extinction of 1 magnitude, either due to a dusty disk or the intervening interstellar medium, would result in J--W2 colors redder by about 0.2--0.3 magnitudes \citep{Cardelli:1989}. As can be seen in Figure~\ref{bd_jw2}, an M5 dwarf, for example, may appear as an M7 or M8 given 1 magnitude of visual extinction towards the star. Given the relative proximity of these stars to Earth, extinctions this high are more likely due to dust within the system. 
Note, however, that 1 magnitude of visual extinction would result in 2.96 magnitudes of extinction at NUV wavelengths \citep{Cardelli:1989}, 
which might, in many or even most cases, lower the apparent UV excess and thereby remove an M-dwarf from consideration as young (given our selection criteria). Thus, in general, we expect GALNYSS sample objects with unusually late inferred spectral types to be dusty systems wherein a disk contributes to the system's mid-IR emission without obscuring the star.

\begin{figure}[htb]
\begin{center}
\includegraphics[width=14cm,angle=0]{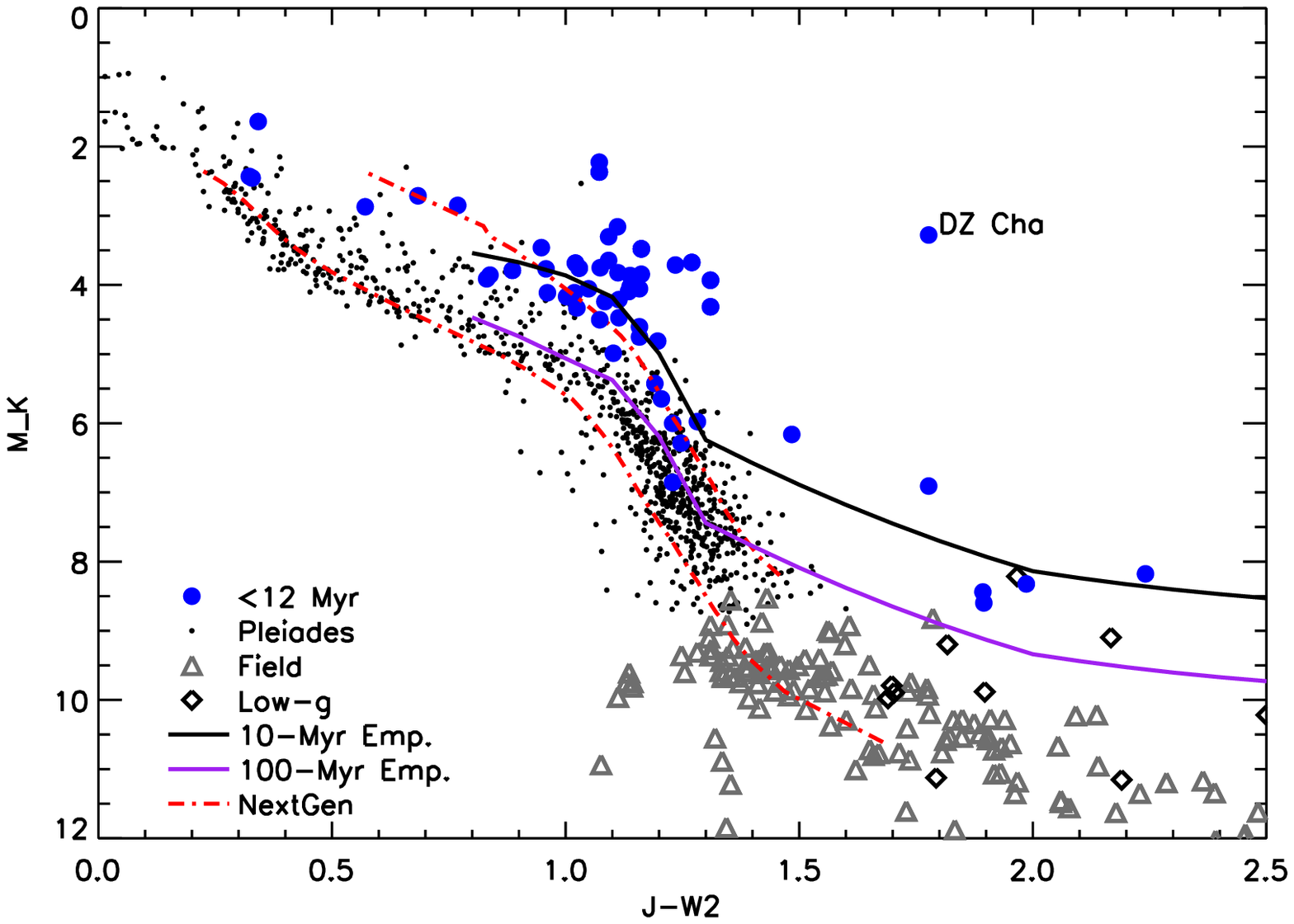}
\end{center}
\caption{Empirical $\sim$10 and $\sim$100-Myr isochrones derived for known young stars listed in 
\citet{Torres:2008} and \citet{Schneider:2012a}. 
NextGen models of the same age are shown for comparison (\citealt{Hauschildt:1999}; see 
{http://phoenix.ens-lyon.fr/Grids/NextGen/}).
Pleiades star data are drawn from \citet{Stauffer:2007}, data for the field (ages $>$100~Myr)
population of low-mass dwarfs are drawn from \citet{Dupuy:2012} and \citet{Faherty:2012}, and data for 
low-gravity objects come from \citet{Faherty:2012}. The red end of the $\sim$10 Myr isochrone (near J--W2 
of 2) is constrained only by TWA~26, 27, 28, and 29, which are M8--M9 dwarfs; see Section~\ref{spectypes} for more 
details. }
\label{fig:iso}
\end{figure}

Given that the ages of our candidates are unknown, we compute distances using two empirical isochrones, one for an age of $\sim$10~Myr and the other for an age of $\sim$100~Myr.
These isochrones are determined from known young, moving group stars (Figure~\ref{fig:iso}). 
Figure~\ref{fig:iso} displays absolute K-band magnitude vs.\ J--W2 color and
shows members of the TWA \citep{Schneider:2012a} and other local groups, with ages $\sim$10~Myr \citep{Torres:2008}, alongside Pleiades candidates from \citet{Stauffer:2007} and cool field dwarfs (ages $>$100~Myr) from \citet{Dupuy:2012} and \citet{Faherty:2012}. 
A piecewise polynomial fit was performed for the $\sim$10~Myr-old stars and subsequently shifted down to match the locus of low-mass Pleiades stars.
For this work we adopt our $\sim$100~Myr-old empirical isochrone given the greater agreement with the low-mass population of the Pleiades and its estimated age of 100--125~Myr (\citealt{Stauffer:2007}; and references therein).
The theoretical 100~Myr NextGen isochrone dips down into the old ($>$100~Myr) population for J--W2$>$1.3.
We note that our empirical $\sim$10~Myr-old isochrone agrees within $\sim$0.5~mag of the NextGen theoretical isochrone over the range 0.8$<$J--W2$<$1.3, corresponding to spectral types $\sim$M0--M5.
A $\sim$0.5~mag error in the absolute magnitude corresponds to a $\sim$20\% uncertainty in the distance.
We thus expect that distances estimated from these isochrones are no more accurate than $\sim$20\%.

Although we extend our isochrones to J--W2 of 2.5, this regime is constrained only by a few TWA members, namely TWA 26, 27, 28, and 29. These are M8--M9 dwarfs with known parallax distances. 
The isochrones are best constrained at early or mid-M spectral types (ie, 0.8$<$J--W2$<$1.3) where more stars have been measured. 
Subsequent work in identifying additional late-M members of these young groups will be key in constraining empirical isochrones and improving on existing theoretical pre-main sequence evolutionary tracks.

\subsection{Kinematic Candidate Selection}\label{uvw}

An important step in confirming membership of candidate stars selected on the basis of their colors and proper motions is determination of Galactic space velocities (UVW). 
We define U as positive towards the Galactic center, V positive in the direction of Galactic rotation, and W positive towards the North Galactic Pole.
Because nearly all candidates lack radial velocities, we use the photometric 10 and 100~Myr distance estimates (as described above) along with the positions and proper motions to estimate UVW with respect to the Sun for a range of assumed radial velocities extending from $-80$ to 80~km/s.
\citet{ZS04} define a ``good UVW box" that contains nearly all young stars within $\sim$100~pc of Earth.
We have used a somewhat broader version of this box extending to more negative U in order to account for members of the Argus Association, all of which have U$\sim$--22~km~s$^{-1}$ \citep{Torres:2008,Zuckerman:2011, Zuckerman:2012}.
If, for the range of radial velocity from $-80$ to $+80$~km~s$^{-1}$, a star has UVW velocities within this acceptable UVW range --- that is, U, V, and W within 0 to $-25$, $-10$ to $-34$, and $+3$ to $-20$~km~s$^{-1}$, respectively --- then it is flagged for followup investigation. 

When calculating UVW velocities, we consider three estimates of proper motions: PPMXL, UCAC4, and our WISE/2MASS estimates. We calculate UVW based on each of these three PM lists individually and then retain only those that lie within the acceptable range of UVW for young stars. 
For approximately one-third of the sample the PMs from all three lists yield a range of acceptable UVW; for another third, two of the three lists yield acceptable UVW; and for the remaining third, only one list yields some UVW entries in the acceptable UVW range. 
Given that nearly all sources have proper motions in PPMXL, we favor PPMXL proper motions in preference to those of UCAC4 and our own (WISE/2MASS-based) estimates.
The UVW estimation is performed via custom IDL routines, but an online Javascript calculator\footnote{For the UVW Javascript Calculator, see \url{http://www.astro.ucla.edu/$\sim$drodrigu/UVWCalc.html}} is available to quickly calculate the potential range of UVW velocities for a single star by varying radial velocity or distance choices.

We tested the robustness of this analysis with known young moving group members listed in \citet{Torres:2008}. 
For objects with spectral types K5 or later, we calculated distances using our isochrones, gathered proper motion information, and calculated UVW velocities as described above. 
We find that 92.5\% of the \citet{Torres:2008} young moving group members are recovered. 
The resulting $\sim$10- and $\sim$100-Myr isochrone distances calculated by our methodology are within 20--30\% of the parallax and kinematic distances listed in \citet{Torres:2008}.

\subsection{Resulting Candidate Nearby, Young M-dwarfs}\label{candidates}

The steps described in Sections~\ref{method:uv}--\ref{uvw} above were performed for the entire GALEX GR~4/5 database, resulting in a list of 32,412 sources (use of the GR~6 database, which was released subsequent to our GR 4/5 database search, adds only 2\% of new sky coverage).
We trimmed the source list by requiring an object's UVW velocities be within the acceptable UVW box for a radial velocity range of at least 15~km/s using the $\sim$10~Myr isochrone distance --- that is, at least $\sim$10\% of the full radial velocity range we tested yields UVW velocities consistent with young-star status, given the star's adopted PM and assuming an age of 10~Myr.
We furthermore required that these sources have $\sim$10~Myr distance estimates that place them within 150~pc of Earth.
After applying these additional constraints, we obtain a list of 2107 candidate nearby, young, low-mass stars.

Despite the greater GALEX coverage in the Northern Hemisphere ($\sim$65\% vs $\sim$57\% in the South), more of our $\sim$2100 GALEX-selected nearby, young, low-mass stars lie in the southern sky (58\% vs 42\% in the North). 
These results are consistent with a preponderance of young moving groups in the South as has been noted previously by \citet{ZS04} and \citet{Torres:2008}.
Given our identification of $\sim$850 Northern young star candidates, the present work may lead to the identification of potential new young moving groups in the Northern Hemisphere.

With Equation~2, we estimate spectral types for all 2107 candidates. 
One surprising result is an apparent population of UV-bright sources with very late spectral types extending into the L-dwarf regime. 
As noted in Section~\ref{spectypes} and further demonstrated in Section~\ref{specoverview}, infrared excesses due to warm dust can produce redder J--W2 colors than expected, and thus Equation~2 may misclassify dusty systems toward later spectral types.
Examining these late-type candidates visually and with SIMBAD revealed that blended sources and carbon stars are also potential sources of contamination. 
However, the dominant contaminant for these late-type objects are galaxies. 
Such contaminants tend to appear around RA and Dec coordinates of approximately (280, 30) and (100, --30) degrees, corresponding to the solar apex and antapex. These regions require low proper motions and moderately large negative or positive radial velocities ($\pm$(20--40)~km/s) in order to match the good UVW box described in Section~\ref{uvw}.
We have visually inspected the list of 87 potentially misclassified L-dwarfs and removed 55 galaxy contaminants and 8 blended sources. In addition, we have cross correlated against SIMBAD and removed an additional 13 known galaxies from the full table.
The final result is a table of 2031 candidate low-mass stars, with perhaps a few percent contamination from other types of objects. 

\begin{figure}[htb]
\begin{center}
\includegraphics[width=14cm,angle=0]{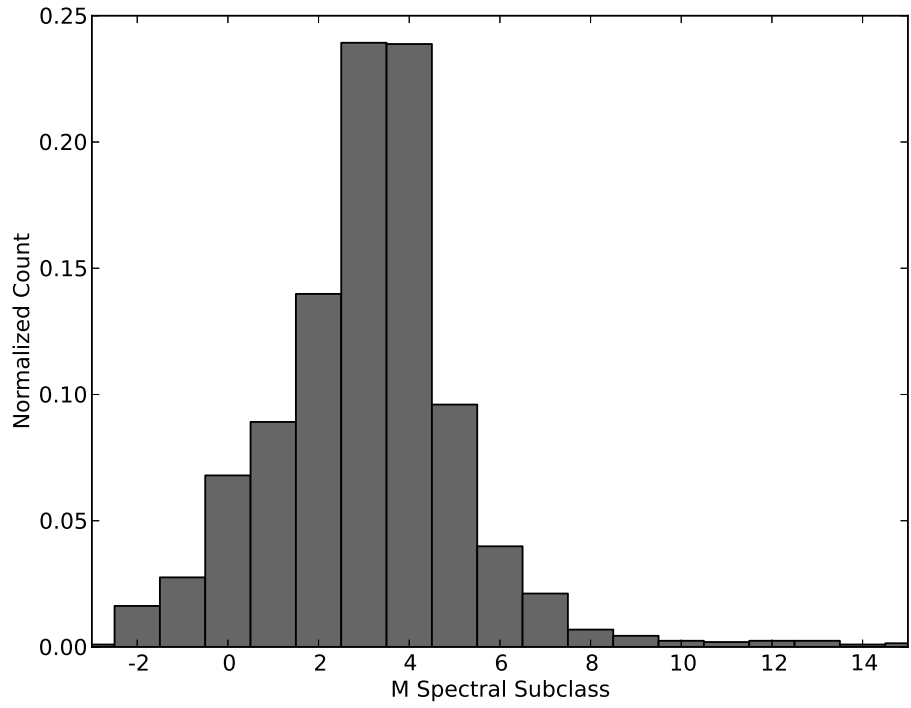}
\end{center}
\caption{Distribution of photometrically estimated spectral types for our sample of 2031 candidates. The numbers on the abscissa indicate the class past M0, so 0=M0, 10=L0, and so forth. We interpret --1 and --2 as indicating spectral types of K7 and K5, respectively (see Sections~\ref{spectypes} and \ref{candidates} for details).
}
\label{specdist}
\end{figure}

The distribution of spectral types is shown in Figure~\ref{specdist}. 
About half our candidates have spectral types between M3 and M4, which is similar to the observed distribution of stars in the immediate solar neighborhood (ie, D$\leq$20~pc; see Figure~7 in \citealt{Farihi:2005}, also \citealt{Henry:2006,Reid:2007,Bochanski:2010,Stelzer:2013}; and references therein).
This suggests that our list of 2031 nearby young star candidates likely includes the missing M3--M5 members of nearby, young moving groups.
Further work will be necessary to see if any bonafide L-dwarfs or dusty M-type stars are also present in our sample.

Rather than publishing the present, full list of 2031 GALNYSS candidates, the vast majority of which remain unconfirmed, we have elected here to focus on a subset of this list (Section~\ref{tuchor}), to demonstrate the basic characteristics of the objects we are identifying. The full GALNYSS table will be presented in a future publication. 

\subsection{Spectroscopic Followup}\label{specfollowup}

In order to confirm that the candidates identified via the methods of Sections~\ref{method:uv}--\ref{candidates} are in fact young low-mass stars, we must follow up with spectroscopic observations designed to exploit diagnostics of youth.
Candidates considered for spectroscopic followup in this and forthcoming papers are those with a broad range of radial velocities that would give UVW velocities consistent with nearby, young star status. 

Low to medium-resolution spectroscopy can be used to confirm spectral types and identify features of youth.
Emission lines from hydrogen and helium can indicate stellar activity common to young stars \citep{White:2003}, while the strength of Na~I absorption lines can indicate low surface gravity (\citealt{Lawson:2009,Schlieder:2012}; and references therein) and, thus, pre-main sequence star status.
Lithium absorption at 6708\AA~is also a widely used diagnostic for youth.
However, lithium has been shown to be rapidly depleted within $\sim$10~Myr or so for mid-M stars \citep{Song:2002,ZS04,Yee:2010}.
Thus the absence of Li absorption only demonstrates that a mid-M star is older than $\sim$10~Myr.
At infrared wavelengths, a triangular (or ``peaky") shape to the H-band profile and weak FeH, Na~I, and K~I \ features are all indicative of low surface gravity \citep{Lucas:2001,Allers:2007,Rice:2010a,Faherty:2013}.
With high spectral resolution, radial velocities can be measured for our candidate stars. The radial velocity can then be used to constrain the UVW velocity so as to solidify young star status and perhaps place the system among known moving groups.

\section{GALNYSS Identification of Nearby, Young, Low-Mass stars in the Tucana-Horologium Association}\label{tuchor}

The Tucana-Horologium Association (Tuc-Hor) was independently identified by \citet{Zuckerman:2000} and \citet{Torres:2000}, originally as two separate nearby moving groups. \citet{Zuckerman:2001} later suggested that these two groups be merged into a single group located $\sim$50~pc from the Earth.
Subsequent studies (see \citealt{Zuckerman:2011}, \citealt{Zuckerman:2012}, and references therein) have proposed additional stars as members of the group. 
\citet{Malo:2012} list 44 bona fide members (of a possible 67) with distances ranging from 36--71~pc.
An age of 10--40~Myr \citep{Zuckerman:2000} and $\sim$30~Myr \citep{Torres:2000} has been proposed for Tuc-Hor members based on the strength of H$\alpha$ emission, Li abundance, X-ray emission, and stellar isochrones (see also Section~2.3.1 in \citealt{Malo:2012}).
For purposes of this paper, we have used the 44 Tuc-Hor members listed in \citet{Torres:2008} and have adopted an age of  $\sim$30~Myr and group distance of $\sim$50~pc.

We have applied the methods described in Section~\ref{fvalue} to our list of $\sim$2000 GALNYSS-identified nearby young star candidates, and have identified a list of 58 candidates in the vicinity of Tuc-Hor whose kinematics are consistent with membership in this association. 
We note, however, that given the similarity in the UVW of members of Tuc-Hor and the $\sim$30~Myr-old Columba Association \citep{Torres:2008} some of these candidates may instead be members of Columba (Section~\ref{banyan}).

\subsection{F-Value Analysis}\label{fvalue}

The large number of candidate nearby, young stars yielded by GALNYSS complicates searches for numerically sparse but kinematically coherent groups of candidates in specific search regions such as Tuc-Hor.
Hence, we devised a method to group together objects with similar proper motions and distance estimates. This involves the calculation of a quantity, which we call $F$, for a given star among a sample of stars in some region of the sky:
\begin{eqnarray*}
F = \sqrt{ (\frac{\mu_{ra}}{\mu_{ra,0}}-1)^2 + (\frac{\mu_{dec}}{\mu_{dec,0}}-1)^2 + (\frac{d}{d_0}-1)^2 } 
\end{eqnarray*}
The ($\mu_{ra,0}, \mu_{dec,0}, d_0$) terms correspond to the averages in proper motion and distance from Earth of the group of stars being compared. For simplicity, uncertainties in these terms are not folded into this expression.
Stars can be either members of a known group, such as the TW~Hya Association, or all GALNYSS sources in a selected area of the sky.
Objects within a chosen threshold F-value are retained as candidate group members and considered for followup study.
For applications of the F-value analysis to an arbitrary region of the sky (i.e., not using an already known moving group to provide the average terms), we perform the analysis in an iterative fashion by selecting a threshold $F_T$ and rejecting 
sources that have F-values larger than adopted cutoffs of $2.5 \times F_T$, $1.25 \times F_T$, and $F_T$, 
and then recomputing the average ($\mu_{ra,0}, \mu_{dec,0}, d_0$) terms for each step using sources that were not rejected.

To validate this technique we applied the F-value analysis to members of the TW~Hya Association (TWA), to Tuc-Hor, and to random fields in the GALNYSS candidate table.
For the 26 TWA members listed in \citet{Schneider:2012a} with measured proper motions, we find average parameters and standard deviations of ($\mu_{ra}=-78 \pm 22$~mas/yr, $\mu_{dec}=-24 \pm 9$~mas/yr, $d=53 \pm 17$~pc) and F-values ranging from above 0 up to 1.2. The two largest F-values (1.1 and 1.2) are found for TWA~12 and 31, both of which have been considered questionable members of the TWA (see \citealt{Schneider:2012a}). The remaining objects have F-values smaller than unity. Tuc-Hor, on the other hand, shows a much broader F-value distribution, with values up to 2.2.
The average terms for the members listed in \citet{Torres:2008} are ($\mu_{ra}=73 \pm 24$~mas/yr, $\mu_{dec}=-36 \pm 32$~mas/yr, $d=49 \pm 9$~pc). 

In contrast, when examining random groups of GALEX young star candidates, we find much larger F-values (in some cases, F$\sim$100). 
This suggests that F-value thresholds of a few are adequate to select objects that may be members of young groups like TWA and Tuc-Hor.
We note that the F-value technique will not work efficiently in well-dispersed stellar moving groups, such as the $\beta$~Pic moving group, where members possess a wide range of proper motions and distances from Earth or when members lie near the location of a group's convergent point. However, it can still be applied in small regions for such groups.

\begin{figure}[htb]
\begin{center}
\includegraphics[width=14cm,angle=0]{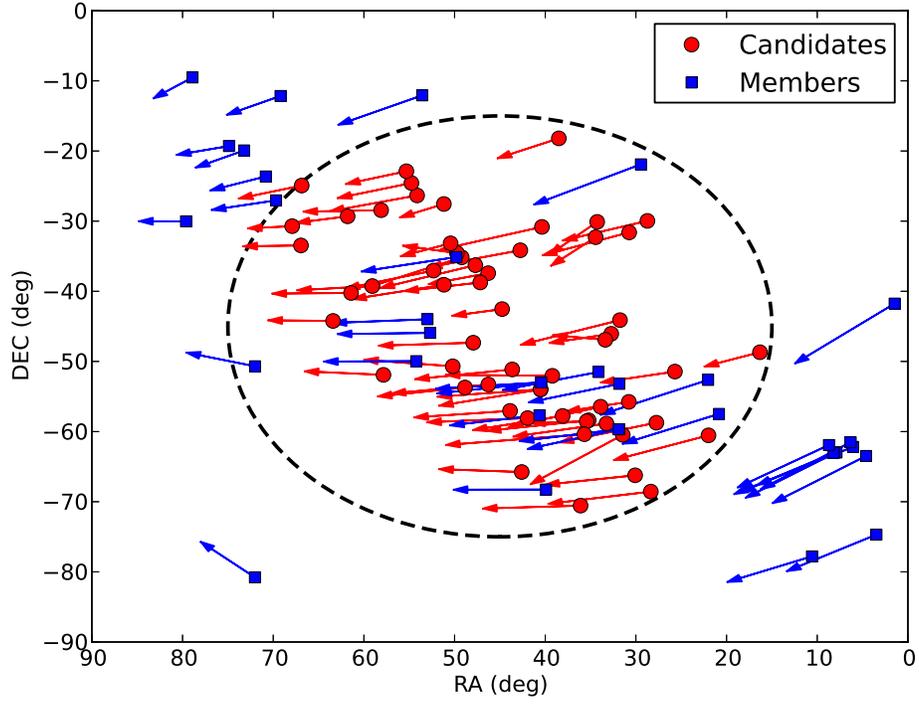}
\end{center}
\caption{Proper motions for the 58 young, low-mass candidates (see Table~\ref{table:th1}) compared with those of known Tuc-Hor members \citep{Torres:2008}. The dashed line highlights the region considered in this study (see Section~\ref{fvalue}).}
\label{th:pm}
\end{figure}

We applied this F-value technique to the 2031 candidate nearby young stars (Section~\ref{candidates}) to investigate the potential presence of young groups of stars at random locations across the sky. 
Candidate groups are generated by selecting starts that lie within 20--30 degrees of a chosen location and applying the iterative method described above.
One of the largest groups so identified is coincident with the location of most known Tuc-Hor members. This group of 58 stars (listed in Table~\ref{table:th1}) was identified considering a search radius of 30 degrees centered at (45, --45) degrees.
The mean proper motions of the stars in Table~\ref{table:th1} ($73 \pm 22$, $-14 \pm 13$ mas/yr) match those of known Tuc-Hor members (see above) reasonably well (Figure~\ref{th:pm}), despite our exclusion of known Tuc-Hor members to determine these values of  $\mu_{ra,0}$, $\mu_{dec,0}$, and $d_0$. 
The F-values listed in Table 2 range up to 2, similar to those of known Tuc-Hor members.
Indeed the distributions of F-values for candidates and for known members are indistinguishable from each other, as shown by a KS~test ($P=0.79$). In contrast, comparing the distributions of either Table~\ref{table:th1} Tuc-Hor candidates or known Tuc-Hor members with that of TWA members yields KS~test $P$= few$\times10^{-3}$, consistent with different underlying F-value distributions for TWA and Tuc-Hor. 
Most Table~\ref{table:th1} candidates, if at the $\sim$30-Myr age of Tuc-Hor, have distances close to $\sim$50~pc, in agreement with the known Tuc-Hor members.
We stress that most of these objects are only candidates at this stage in our work. 
Furthermore, additional Tuc-Hor candidates likely exist beyond the region we have searched and are not considered in this paper.
In Sections~\ref{specoverview}--\ref{note2}, we discuss properties of the Table~\ref{table:th1} candidates, and in fact are able to rule out Tuc-Hor membership for some of them. 
Note that, although these rejected candidates are unlikely to be Tuc-Hor members, they may be members of other moving groups (e.g., $\beta$~Pic, Columba, or AB Dor) and, hence, may still be young (age $<100$ Myr) stars.

\subsection{Convergent Point Analysis}\label{cpanalysis}

To complement the F-Value analysis described in Section~\ref{fvalue}, we also performed a convergent point analysis on the candidate Tuc-Hor members. 
The convergent point analysis methodology has been described in detail elsewhere \citep{Mamajek:2005,deBruijne:1999,Jones:1971}; here, we have applied a simplified algorithm. 
This analysis consists of selecting a convergent point location and determining the proper motions ($\mu_{\perp}$) perpendicular to the great circle connecting a star to the convergent point. The typical analysis removes suggested but discrepant members from a list until the convergent point analysis passes some threshold. We have simplified the procedure by neglecting this step; that is, we retain potentially discrepant members while determining the location of the convergent point. This simplification, combined with the coarse, 1-degree grid we use for convergence point calculations, likely results in a convergent point that is less accurate than would be obtained for a more rigorous treatment in which outliers are rejected. However, the simplified algorithm is sufficient for the purposes of this work, namely, identifying candidate new members of moving groups.

Given a convergent point location for a moving group and the perpendicular motion of a candidate group member, one can estimate a membership probability:
\begin{eqnarray*}
P = \text{exp}(-\frac{1}{2} \frac{\mu_{\perp}^2}{(\sigma_{\perp}^2 + \sigma_{int}^2)})
\end{eqnarray*}
The quantity $\sigma_{\perp}$ incorporates error terms in the proper motion and convergent point location (Equation~12 in \citealt{deBruijne:1999}), while $\sigma_{int}$ accounts for the internal dispersion in the group. 
Given our coarse determination, we adopt an uncertainty of 2~degrees for the convergent point location. 
While the quantitative probabilities so derived may be approximate, they nevertheless present a qualitative estimate of membership probability given an object's proper motion and the estimated convergent point location of a group.

We have applied this convergent point analysis to the known members of Tuc-Hor listed in \citet{Torres:2008}.
We determine a convergent point of (119, --27) degrees for Tuc-Hor and used
proper motions, distances (both parallax and kinematic; \citealt{Torres:2008}), and radial velocities \citep{Torres:2006} to determine a mean total heliocentric space velocity magnitude of $23.3\pm1.3$~km/s (Table~\ref{cpinfo}); here the quoted error is the standard deviation of the velocities.
We thus adopt an internal velocity dispersion ($\sigma_{int}$) of 1~km/s and, given the known member distance distribution, a group distance of 50~pc in our analysis. 
The group distance is used to convert the velocity dispersion in km/s to mas/yr (with the result 4 mas/yr for Tuc-Hor).
However, the convergent point location is not very sensitive to these estimates (see \citealt{Mamajek:2005}).

The space velocity of known members can be combined with the angular distance between a given target star and the convergent point to estimate the radial velocity and distance the target would have if it were a member. 
We have created an online Javascript calculator for this purpose\footnote{The online calculator takes, as inputs, moving group convergence point data and stellar coordinates and proper motions, and outputs the membership probability as well as expected kinematic distance and radial velocity the star would have as a member of the group of interest; see \url{http://www.das.uchile.cl/$\sim$drodrigu/CPCalc.html}}.
Table~\ref{cpinfo} lists convergent point information for various groups determined via the method described above. These can be used as inputs for our Javascript tool.

\begin{figure}[htb]
\begin{center}
\includegraphics[width=14cm,angle=0]{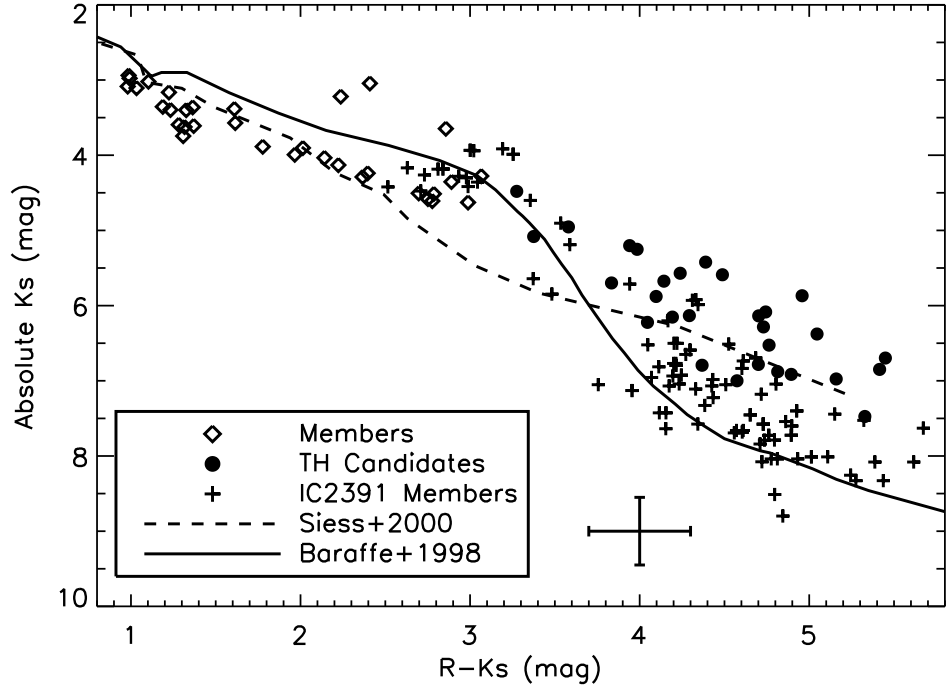}
\end{center}
\caption{Ks vs R--Ks color-magnitude diagram for previously established Tuc-Hor members and the Table~\ref{table:th1} candidate objects. 
Ks magnitudes come from 2MASS; R comes from NOMAD, which in turn are drawn from either USNO-B1 or UCAC2. Given that the R magnitudes come from separate catalogs, we adopt a 0.3 magnitude error for R--Ks.
This is shown as the approximate typical uncertainty (error bar at bottom center), which includes a 20\% distance error.
For the candidate objects, we use kinematic distances derived in our convergent point analysis (see Section~\ref{cpanalysis}). We do not show candidates whose properties indicate they may not be Tuc-Hor members (see Section~\ref{banyan}, Appendix~\ref{notesys}, and Table~\ref{tab:final}).
For comparison, we show members of the $\sim$50~Myr-old cluster IC2391, located 155~pc from Earth \citep{Barrado:2001}.
Theoretical isochrones for age 30~Myr from \citet{Baraffe:1998} and \citet{Siess:2000} are also shown. 
The 3 Tuc-Hor members above the sequence with R--K between 2 and 3 are HD~3221, BD--20~951, and CD--53~544.
BD-20~951 is a known spectroscopic binary, while CD--53~544 is a visual double \citep{Torres:2008}.
}
\label{th:cmd}
\end{figure}

The kinematic distance and radial velocity estimates for the 58 Tuc-Hor candidates (Table~\ref{table:th1}) are tabulated, along with the probabilities described here, in Table~\ref{table:th2}.
In Figure~\ref{th:cmd}, we use our kinematic distances to construct a color-magnitude diagram for the candidate Tuc-Hor stars. 
The Tuc-Hor low-mass candidates lie above the IC2391 members in the color-magnitude diagram, suggesting the Tuc-Hor candidates have ages $<$50 Myr, consistent with the age of Tuc-Hor as inferred from prior studies (see Section~\ref{tuchor}).
A pair of theoretical 30~Myr-old isochrones are also shown \citep{Baraffe:1998, Siess:2000}; the agreement is marginally better with the \citet{Siess:2000} isochrones for the Tuc-Hor candidates. 
When compared to the absolute magnitudes of field population M-dwarfs (e.g., \citealt{Kraus:2007}), we find that our candidates, when located at the kinematic distances we estimate, are all over-luminous, as expected for young systems.

\subsection{Spectroscopic Observations} \label{specoverview}
\subsubsection{WiFeS}\label{wifes}

To investigate the nature of the Tuc-Hor candidates, we used the WiFeS spectrograph at the Siding Springs Observatory 2.3-m telescope to observe some of these systems at resolution $\sim$3000 or $\sim$7000. 
WiFeS \citep{Dopita:2007} is a double-beam, image-slicing integral field spectrograph, and provides a 25\arcsec$\times$38\arcsec\, field with 0.5\arcsec\,pixels.
Observations over the spectral range 3200--9800\AA\, were carried out in 2012~September and October.
Figure~\ref{wifes} shows the spectra, covering the H$\alpha$ and Li region, for stars observed thus far. Initial analysis has focused on measurements of H$\alpha$ emission, lithium absorption at 6707\AA, and sodium absorption at 8200\AA. To derive accurate spectral types we computed the TiO5 index as described in \citet{Reid:1995}. 
We list our measurements in Table~\ref{meas1}, including a handful of cases in which published spectroscopy exists in the literature.

Figure~\ref{sptype} compares the spectral types initially estimated with J--W2 colors to those derived from the TiO5 index (accurate to $\pm$0.5) for Tuc-Hor candidates and young low-mass candidates from \citet{Rodriguez:2011}. 
There is evidently sufficiently good agreement between these spectral type estimation methods that a disagreement --- in the sense that J--W2 is too red --- likely indates the presence of emission from warm dust grains.
Such appears the case for J0215--56, J0259--42, and J0324--27 (see Figure~\ref{sptype}).
We note that at an age of $\sim$30~Myr, brown dwarfs are expected to have spectral types later than $\sim$M6 \citep{Burrows:1997,Baraffe:1998}. 

\subsubsection{SpeX}\label{thspex1}
In addition to optical spectra, near-infrared spectra for J0202--31, J0259--42, and J0324--27 were obtained with the SpeX spectrograph \citep{Rayner:2003} on the 3-m NASA IRTF.
Observations were carried out on 2012 December 25 and 2013 January 1 (UT)  with the spectrograph in prism mode with the 0.5 or 0.8$\arcsec$ slit, 
resulting in $R = \lambda / \Delta \lambda \approx$ 100 over the wavelength range 0.7--2.5$\mu$m (Figure~\ref{th_irtf}). 
Observations were carried out with an ABBA dither pattern along the slit, with exposure times determined by the target magnitudes.
A0V stars observed immediately after each target were used as standards for flux calibration and telluric correction. 
All data were reduced with SpeXtool version 3.3 \citep{Vacca:2003,Cushing:2004} and standard settings.

At the IRTF, we resolved J0202--31 as a $\sim$0.5$\arcsec$ binary; while guiding, the slit switched between both stars. 
Comparison with field dwarf standards shows the spectrum of the pair to be an M4 type, in agreement with the spectral type determined from the optical spectrum --- which did not resolve the system --- suggesting J0202--31 is an equal mass binary.
The SpeX spectrum of J0259--42, which displays a very red J--W2 color, indicates a spectral type of M5, also in agreement with the optical spectrum.
We do not have an optical spectrum for J0324--27, but its SpeX spectrum indicates it is an M5.5, rather than an M9 as indicated by the J--W2 color.
As we discuss in Appendix~\ref{notesys}, both J0259--42 and J0324--27 have WISE infrared excesses that indicate the presence of circumstellar disks. Both show Br-$\gamma$ emission and weak Na~I absorption, which may indicate youth. However, other features, such as the lack of Li absorption in the optical and the shape of the H-band profile (see Section~\ref{specfollowup}), are inconclusive.
J0202--31 does not show any clear signatures of youth in its near-IR spectrum.

\subsubsection{H$\alpha$ and Li}\label{spec1res}

Equivalent widths of H$\alpha$ emission and Li absorption for some of the Tuc-Hor candidates are listed in 
Table~\ref{meas1}. 
Uncertainties for the equivalent width measurements are computed by using the rms scatter from a linear fit to the continuum around each line and propagating this with a Monte Carlo method. For non-detections, we quote only the 1-$\sigma$ uncertainties.
With the exception of J0315--53, no Table~\ref{meas1} star observed with WiFeS shows clear Li absorption at 
6708\AA. This is not surprising as, at the expected $\sim$30~Myr age of these candidates, lithium should 
have been depleted for early to mid M-type stars (see Section~\ref{specfollowup}). 
J0315--53 is therefore a noteworthy exception as it has a spectral type near M5 and Li absorption EW of $\sim$400m\AA. One possibility is that J0315--53 is a member of the $\sim$10~Myr $\beta$~Pictoris moving group. To investigate this possibility, we carried out a convergent point analysis using $\beta$~Pic members in a similar fashion to those of Tuc-Hor (Section~\ref{cpanalysis}). 
However, with our estimated convergent point for the $\beta$~Pic moving group (Table~\ref{cpinfo}), we find a probability of only 6\% that J0315--53 is a member of $\beta$~Pictoris (similar to the 0.3\% probability returned by the Bayesian analysis of \citealt{Malo:2012}). 
Hence, while the lithium absorption EW may suggest an age comparable to the $\beta$~Pic moving group, J0315--53's kinematics nevertheless suggest it is a strong candidate for membership in Tuc-Hor.
A handful of Table~\ref{table:th1} systems have published spectroscopy available in 
\citet{Riaz:2006} or \citet{Torres:2006}, including J0236--52, another system with clear Li absorption. 
J0315--53 and J0236--52, among other systems, are considered in more detail in Appendix~\ref{notesys}.

All Table~\ref{meas1} stars show H$\alpha$ emission. However, this is not necessarily indicative of youth, as 
H$\alpha$ activity can still be seen among old M-dwarfs \citep{West:2008,West:2004}. \citet{West:2008} 
predict activity lifetimes of order $\sim$1~Gyr for M2 stars and $\sim$6~Gyr for M6 stars. In fact, 
M5--M9 stars in the field are commonly seen to show H$\alpha$ activity, with the latest types nearly 
always showing activity regardless of age \citep{West:2004}. To attempt to determine whether our 
H$\alpha$ detections are indicative of ages $<$1~Gyr, we make a comparison between our sample and the 
older sample studied in \citet{West:2004}. We compute log $L_{H\alpha} / L_{bol}$ for Table 5 stars as described in \citet{Walkowicz:2004}.  Results are 
illustrated in Figure~\ref{halpha} along with the average $L_{H\alpha} / L_{bol}$ from \citet{West:2004}. 
We find that most of the Table~\ref{meas1} sample have larger $L_{H\alpha} / L_{bol}$ than the field population. 
This suggests that the Table~\ref{meas1} stars are statistically younger than the field population studied in 
\citet{West:2004}. 
This is not surprising given that we have selected these candidate Tuc-Hor M-dwarfs based on UV 
emission from GALEX which likely originates from the active chromospheres of young stars (see Introduction). 

\subsubsection{Sodium as an Indicator of Youth}\label{sodium}

Several recent studies have shown that Na~I absorption can be used as a diagnostic of stellar surface gravity and thus of youth \citep{Schlieder:2012,Lawson:2009,Mohanty:2004a,Mohanty:2004b}.
In particular, young objects are expected to have lower surface gravities and hence weaker Na~I absorption. 
The spectral feature used in these studies is the Na~I doublet at 8183 and 8195\AA\, (see Figure~\ref{naspec}). 
\citet{Lawson:2009} construct an index based on the average flux off and on the doublet ($F_{8148-8172} /F_{8176-8200}$), whereas \citet{Schlieder:2012} use the EW summed over both lines. 
We measured both of these quantities for the Tuc-Hor candidates for which we obtained WiFeS spectra and list them in Table~\ref{meas1}.
The Na~I index is measured in spectra that have been re-binned to a resolution of R$\sim$800, to match that used in \citet{Lawson:2009}.

Figure~\ref{naindex} compares the Na~I index with spectral type for Table~\ref{meas1} stars. 
Young low-mass candidates from the \citet{Rodriguez:2011} study of TWA and Scorpius-Centaurus regions are also shown. 
Empirically determined trends for nearby $\sim$10 Myr old moving groups and (old) field dwarfs are also presented \citep{Lawson:2009}, where we have used the relations in \citet{KH:1995} to convert the \citet{Lawson:2009} R--I colors to spectral type. 
The Tuc-Hor candidates have higher Na~I indices than the young (ages $\sim$10--20~Myr) candidates from \citet{Rodriguez:2011}, consistent with the older ages expected for Tuc-Hor members.
While some Tuc-Hor candidates are close to the field dwarf line, these tend to lie in a region where the Na~I index for  young and old stars start to converge (ie, spectral types earlier than about M4). 
The Na~I 8200\AA\, doublet is therefore only useful as an age indicator for stars later than $\sim$M4, as also demonstrated by \citet{Schlieder:2012}.
Figure~\ref{naindex} indicates that most of the Tuc-Hor candidates lie in a region suggestive of an age older than $\sim$10~Myr, but younger than $\sim$1~Gyr.

\subsection{X-ray Emission}\label{xraysec}

Of the 58 Tuc-Hor candidates, only 20 have X-ray detections in the ROSAT All-Sky Survey (RASS). 
Table~\ref{xray1} lists these ROSAT counterparts and their estimated values of $L_X/L_{bol}$. 
To estimate the  $L_X/L_{bol}$ ratios, we convert the X-ray count rates to $F_X$ (ergs cm$^{-2}$s$^{-1}$) using an energy conversion factor (ECF) of $1.25\times10^{11}$ (counts cm$^2$ erg$^{-1}$; \citealt{Neuhaeuser:1995}) and bolometric fluxes calculated using the bolometric corrections listed in \citet{KH:1995}.
{The adopted ECF is similar to that used in prior studies for stars in the same general age range \citep[i.e., 10--30 Myr; e.g.,][and references therein]{Kastner:1997,Rodriguez:2011}.} 
For those stars without X-ray detection, we calculate an upper limit on $L_X/L_{bol}$ by adopting a RASS flux limit of 
$2\times10^{-13}$ ergs~cm$^{-2}$~s$^{-1}$ \citep{Schmitt:1995}. All detections and most upper limits have 
log~$L_X/L_{bol}$ close to --3, the saturation limit for M-dwarfs (e.g., \citealt{Riaz:2006}). 
The one notable exception is J0324--39, with log~$L_X/L_{bol}$ of --2.26. It is thus possible this star may 
have been observed by RASS during a flare event. 
The X-ray detection fraction among our sample (20/58 = 0.34) is similar to that noted in \citet{Rodriguez:2011} for TWA and Scorpius-Centaurus low-mass candidates (14/54 = 0.26). 

Figure~\ref{fig:xray1} compares $L_X/L_{bol}$ vs.\ spectral type for the Table~\ref{table:th1} candidates;
Table~\ref{xray2} summarizes the ROSAT X-ray detection statistics for the sample divided in bins of spectral type and distance. 
Figure~\ref{fig:xray1} and Table~\ref{xray2} indicate that the frequency of RASS X-ray detections strongly depends on spectral type, i.e., the RASS detection frequency drops for stars M4 and later. 
The low rate of X-ray detections among our sample indicates that GALEX observations are capable of detecting active late M-dwarfs that have been missed in the RASS.

\subsection{Comparison with BANYAN} \label{banyan}

\citet{Malo:2012} published a Bayesian analysis tool (BANYAN) to estimate membership probabilities, 
distances, and radial velocities for candidates of known local young moving groups. The output is thus similar to that 
returned by the convergent point method employed here (Section~\ref{cpanalysis}) and serves as a useful comparison. 
In Table~\ref{table:th2}, we list the probabilities, distances, and radial velocities for the 58 Tuc-Hor 
candidates as estimated by both our convergent point analysis tool and that of \citet{Malo:2012}. The two tools return similar predicted radial velocities and kinematic distances, though the membership probabilities differ in detail. 
Comparing Tuc-Hor members with Hipparcos distances \citep{Torres:2008}, we find good agreement between our kinematic distances, the BANYAN distances, and the Hipparcos distances (see Table~\ref{banyan2}). Given this agreement, we adopt a conservative uncertainty of $\sim$20\% for our kinematic distances.

There are several cases where we find a high likelihood of Tuc-Hor membership with our convergent point 
analysis, but BANYAN returns a very low probability. In many of these cases, BANYAN returns a higher 
likelihood for membership in a different group. Furthermore, we find that for those objects with low 
BANYAN Tuc-Hor likelihoods, we predict much larger distances ($\sim$60--120~pc) than the average Tuc-Hor member distances (most of which lie in the range 40--60~pc). 
Figure~\ref{fig:xyz} shows the XYZ coordinates of the Table~\ref{table:th1} candidates, adopting our kinematic distance estimates, and compares these XYZ values to those of known Tuc-Hor members from \citet{Torres:2008}. While most candidates have similar XYZ to those of known members, there are some prominent outliers (see Appendix~\ref{notesys}). 
For these outliers, BANYAN returns low likelihoods for membership in Tuc-Hor.
Hence, the combination of probability and predicted distance in our convergent point analysis is analogous to the probability returned by the BANYAN code. 

\subsection{Notable Systems}\label{note2}

Among Table~\ref{table:th1} stars, there are several systems that are noteworthy for one or more reasons. For example, two systems (J0259--42 and J0324--27) have clear signs of an infrared excess, as first noted in Section~\ref{thspex1} (see also Figure~\ref{fig:sed}), while another system, J0242--53EW, constitutes a wide separation low-mass binary. Other systems have been previously mentioned in the literature; for some systems, we have now measured radial velocities. For systems with radial velocities we calculate UVW velocities and list them in Table~\ref{tab:uvw}. 
Some Table~\ref{table:th1} stars appear to have low likelihoods ($\lesssim$60\%) of Tuc-Hor membership using the convergent point analysis or the Bayesian methodology of \citet{Malo:2012}, but have higher likelihoods for membership in other groups. 
These noteworthy stars are discussed individually in Appendix~\ref{notesys}.
Candidates that remain plausible members of moving groups (after the discussion in Appendix~\ref{notesys}) are summarized in Table~\ref{tab:final}, but we caution that measurements of radial velocities and distances will be necessary to fully confirm or rule out membership.

\section{Summary}

We have carried out an all-sky GALEX-WISE-2MASS search for nearby (D$\lesssim$100 pc), young (age 10--100 Myr), low-mass (mostly M-type) stars. We refer to this search as the GALEX Nearby Young-Star Survey, or GALNYSS. On the basis of their UV/IR colors and space motions, where the latter are inferred from proper motions and photometry, we identify 2031 candidate nearby, young M stars. The spectral type distribution of these candidates peaks near M3-M4. Most prior searches for young stars in nearby moving groups have focused on bright stars or on those with X-ray emission. Hence, these searches have tended to lack the sensitivity to detect the mid-M stars that dominate the list of nearby, young star candidates we have identified via GALNYSS. Given that M3--M5 stars constitute about half of all stars in the solar neighborhood, our survey thereby helps fill an important niche in the study of nearby young moving groups.

We consider a subset of 58 low-mass stars among the larger GALNYSS sample that lie in the vicinity of previously proposed members of the Tucana-Horologium Asssociation.
We have developed and applied a grouping algorithm to select these stars, and we further confirm them as a kinematically coherent group via convergent point and Bayesian analyses \citep{Malo:2012}. 
Optical and near-IR spectroscopy obtained for some of the subsample of 58 Tuc-Hor candidates appear to confirm their young ages ($\sim$30--100 Myr), based on their H emission lines and Li and Na~I absorption line strengths. 
The color-magnitude diagram positions of the candidate Tuc-Hor members (as inferred from their kinematic distances) relative to those of low-mass IC2391 members also indicate that the candidates are likely younger than $\sim$50 Myr.
Only roughly 1/3 of the sample objects are detected in the ROSAT all-sky X-ray survey, indicating that GALEX is capable of identifying young stars that are too faint to have been detected in the most sensitive extant all-sky X-ray survey.

We find that half of the 58 candidates have kinematic and spectroscopic properties consistent with membership in Tuc-Hor (Table~\ref{tab:final}). Many of the other candidates may instead members of other young groups. Specifically, two stars are potential members of the AB Dor Moving group, while 12 may be new Columba Association members. 
Two of the Columba candidates have infrared excesses indicating the presence of warm circumstellar disks in these systems.
While the 15 remaining stars among the group of 58 do not have kinematics that might place them among the known young moving groups, their  UV excesses suggest they nevertheless could be young; these stars warrant  further examination.

Radial velocity measurements of these 58 candidate young, low-mass stars will be required to further confirm membership in Tuc-Hor or in other young moving groups. In subsequent work, we will explore the youth and kinematics of other subgroups among the list of nearby, young star candidates compiled for the GALEX Nearby Young-Star Survey.

\acknowledgements
{\it Acknowledgements.} 
We thank the anonymous referee for useful feedback that strengthened this paper.
This publication makes use of data products from GALEX, operated for NASA by the California Institute of Technology; the Two Micron All Sky Survey, which is a joint project of the University of Massachusetts and the Infrared Processing and Analysis Center/California Institute of Technology, funded by the National Aeronautics and Space Administration and the National Science Foundation; and the Wide-field Infrared Survey Explorer, which is a joint project of the University of California, Los Angeles, and the Jet Propulsion Laboratory/California Institute of Technology, funded by the National Aeronautics and Space Administration.
This work has used the SIMBAD database, operated at CDS, Strasbourg, France.
This research was supported in part by NASA Astrophysics Data Analysis Program grant NNX12AH37G to RIT and UCLA. 
D.R.R. acknowledges support from project BASAL PFB-06 of CONICYT, a Joint Committee ESO-Government of Chile grant, and FONDECYT grant \#3130520.

\clearpage

\appendix

\section{Notes on Individual Systems}\label{notesys}

{\it J0202--31:} This M4 system has a higher likelihood of belonging to the Columba Association (BANYAN: 74\%, Convergent point: 90\%) than to Tuc-Hor (BANYAN: $\sim$20\%, Convergent point: 61\%). 
While the kinematics favor Columba membership, the weak H$\alpha$ emission and strong Na~I absorption instead suggest the system may be old.
With an R$\sim$7000 WiFeS spectrum, we estimate a radial velocity of $16.7\pm1.5$~km/s and list calculated UVW velocities in Table~\ref{tab:uvw}.
The UVW are a decent match to those listed in \citet{Torres:2008} for Columba. Furthermore, the XYZ position (--10, --13, --58~pc) is also consistent with the XYZ of Columba members listed in \citet{Torres:2008}.
Adopting an age 30 Myr (for Columba) yields reasonable agreement with the predicted kinematic distance (60~pc) assuming the object is an equal-components binary, as revealed by IRTF imaging (see Section~\ref{thspex1}).
However, our spectroscopic observations of this system suggest an H$\alpha$ luminosity consistent with that seen in old active stars (see Figure~\ref{halpha}) and likewise Na~I 8200\AA\ absorption that is stronger than expected for young stars (see Figure~\ref{naindex}). 
The Na~I doublet at 2.2$\mu$m has equivalent width of 4.8\AA, comparable in strength to that seen in a field M4 dwarf ($\sim$4\AA). In comparison, other candidate Columba members of spectral type M5 have weaker Na~I (J0259--42 and J0324--27:\ $\sim$3\AA) when compared to the field (M5:\ $\sim$7\AA). 
J0202--31, along with J0259--42 and J0324--27, has not been detected in X-rays, though this could be a combination of the spectral types (M4--5) and distances (60--100~pc) for these candidates.
Hence, we conservatively infer that J0202--31 is not likely to be a member of Columba. 

% 60% AB Dor:
{\it J0203--55:} An M4.5 system with moderate BANYAN likelihood of membership in AB~Dor (69\%). Our convergent point, however, predicts high likelihoods (99\%) for either Tuc-Hor, Columba, or AB~Dor. The radial velocity required for membership for these groups is 9, 13, and 25~km/s, respectively.
However, the predicted distances for Tuc-Hor membership ($\sim$80~pc) is too large compared to that of known members. Our $\sim$100~Myr isochrone predicts a distance of 50~pc, which is moderately close to the kinematic distance as an AB~Dor member ($\sim$70~pc) given the 20\% uncertainties. 
We tentatively place J0203--55 as an AB~Dor candidate in Table~\ref{tab:final}, but caution that, as with most targets in this study, further work will be necessary to fully confirm membership.

{\it J0210--46:} This low-mass star forms a visual binary (21.7\arcsec\, separation) with CD-46~644, a known 
AB~Dor member \citep{Torres:2008}. Although we removed young stars from \citet{Torres:2008} as part of 
our analysis, this low-mass companion is not listed in the main AB~Dor member table (see Table~13 in 
\citealt{Torres:2008}, but see their Table~14) and was thus recovered here. 
Curiously, the UCAC4 proper motions we've used yield higher likelihoods ($>$90\%) of membership in Columba, in contrast to other proper motion estimates which place it in AB~Dor.
If bound, the CD-46~644 and J0210--46 system would have a separation of $a\sim$1500~AU and binding energy of $-E=GM_1M_2/a\approx15\times10^{41}$ ergs, assuming masses of 0.85 and 0.15$M_\odot$. This is similar to the binding energies of known low-mass binaries; see discussion of J0242--53EW, below.
Given the proximity to a previously proposed AB~Dor member and possibility that it is bound, in addition to prior mention in the literature \citep{Torres:2008}, we place J0210--46 as a plausible AB~Dor member in Table~\ref{tab:final}.

{\it J0220--58:} This M3 system has a kinematic distance estimate of 48~pc and we measured a radial velocity of $7.4\pm1.5$ with WiFeS. 
The UVW velocities for J0220--58, listed in Table~\ref{tab:uvw}, match those of known Tuc-Hor members very well (see also Figure~\ref{fig:uvw}). While the R$\sim$7000 WiFeS spectrum does not cover the Na~I 8200\AA\, doublet, the H$\alpha$ emission and lack of Li absorption is typical of other Tuc-Hor candidates and suggests a youthful age, but $>$10~Myr. The spectroscopic and kinematic properties suggest J0220--58 is a new low-mass member of Tuc-Hor.

{\it J0236--52:} This system, also designated as GSC 8056--0482, is an M2 star first considered a member of Tuc-Hor in \citet{Torres:2000}. 
However, both \citet{Torres:2000} and \cite{ZS04} warn that its large lithium 
EW ($\sim$300\AA) is inconsistent with Tuc-Hor membership. Such a large EW is 
indicative of a system younger than Tuc-Hor, a situation similar to that of J0315--53. We estimate a 
convergent point distance of $42\pm8$~pc, larger than previous photometric distance estimates of 
$\sim$25~pc \citep{Torres:2000,ZS04}. These previous distance estimates are based on 30~Myr-old 
isochrones, assuming the star is single. We note that \citet{Chauvin:2010} rule out the presence of any 
close, low-mass companions on the basis of adaptive optics imaging. The discrepancy between the 
convergent point distance and the isochrone distance may be the result of uncertainties in the spectral 
type of the star, the isochrone models, or the convergent point estimate itself (although this 
seems unlikely, given the close agreement of convergent point distances with the parallax distances of 
known Tuc-Hor members; see Table~\ref{banyan2}). Another possibility is that the system is younger and 
hence more distant than expected, which would be more consistent with the measured lithium abundance. 
We note that our 10~Myr isochrone suggests a distance of 45~pc for this system, which implies J0236--52 should be closer to the Earth if its age were 30~Myr.
While the BANYAN tool returns a 60\% likelihood of Tuc-Hor membership, our convergent point analysis returns effectively 0\%.

In Table~\ref{tab:uvw} we list calculated UVW velocities using a previously measured radial velocity ($16\pm1$~km/s; \citealt{Torres:2006}) and distance estimates of 42 and 25~pc.
In Figure~\ref{fig:uvw} we plot the UVW velocities for this star for the 42 pc distance estimate. 
While UVW velocities for the 42 pc distance are not a perfect match to those of Tuc-Hor, given the large uncertainties, J0236--52 may still be a member. 
At a distance of 25~pc, the UVW velocities appear more consistent with the $\sim$10~Myr-old $\beta$~Pic moving group.
While the Li absorption suggests a young age, comparable to that of $\beta$~Pic, a parallax distance is essential if we are to place this system among the nearby moving groups. 
Given the conflicting results for this system, J0236--52 is not listed as a plausible Tuc-Hor member in Table~\ref{tab:final}.

{\it J0242--53EW:} This system consists of a pair of M4.5 stars separated by $\sim$15\arcsec. 
We label the individual systems within the pair as the East (J0242--53E) or West (J0242--53W) component.
Both show high likelihoods of membership in Tuc-Hor with a kinematic distance of 44~pc. At that distance, the pair of stars would be separated by $\sim$700~AU. A single X-ray source is detected with ROSAT. While we only have spectroscopy for J0242--53W, the spectra are consistent with that of a $>$10~Myr dwarf.
Membership in Tuc-Hor remains to be confirmed, but this represents a new wide-separation low-mass binary.
Assuming a mass of $\sim$0.1~$M_\odot$ for each component, we estimate a binding energy of $2.5\times10^{41}$~ergs. 
While a minimum binding energy of $\sim20\times10^{41}$~ergs is typically cited for tight, low-mass binaries \citep{Close:2007,Close:2003,Burgasser:2003}, there is a growing population of systems with smaller binding energies comparable to or lower than that of J0242--53EW (e.g., see Figure~8 in \citealt{Faherty:2011}).

{\it J0254--51:} This system, also designated as GSC 8057--0342, is an M1.5 star first identified by 
\citet{Torres:2000} as a possible member of Tuc-Hor (referred to as the Horologium Association in that 
paper). Lack of distance and radial velocity information prevented membership confirmation at the time. 
The re-identification of this candidate suggests future work is worthwhile to verify its membership 
status.

{\it J0259--42:} This system is one of only two candidates in our sample that display very red WISE colors 
(see Figure~\ref{fig:sed}). \citet{Schneider:2012b} note that $W1-W4$ colors $>$1 among M-type systems 
are indicative of mid-IR excess and interpret this as the presence of a dusty circumstellar disk.  
The $W1-W4$ color for J0259--42 is 3.6, an excess comparable to that seen in some TW~Hya members such 
as TWA 31 and 32 \citep{Schneider:2012a}.  Only very young M-type stars show such mid-IR excess emission 
\citep{Schneider:2012b}.  
On the other hand, lack of Li absorption suggests an age $\gtrsim$10~Myr. 
In terms of Na~I absorption, J0259--42 is similar to the other Table~\ref{table:th1} Tuc-Hor candidates, suggesting a similar age. 
Near-IR spectroscopy of J0259--42 (see Figure~\ref{th_irtf}) reveals some signatures of youth, such as Br-$\gamma$ emission 
and weaker Na~I absorption than the field, but the distinctive triangular H-band profile seen in 
low-surface-gravity dwarfs is lacking. However, our kinematic distance estimate (108~pc) places it too far away for membership in Tuc-Hor. 
If J0259--42 were $\sim$10~Myr-old, it would have a distance of $\sim$140~pc; 
in contrast, as an old field dwarf, it would lie only $\sim$40~pc away \citep{Kraus:2007}. 

The radial velocity measured for J0259--42 with an R$\sim$7000 WiFeS spectrum is $15.3\pm1.5$~km/s. The convergent point (Section~\ref{cpanalysis}) and Bayesian \citep{Malo:2012} methods both predict radial velocities of $\sim$17~km/s for membership in the Columba Association, with kinematic distances of 106 and 92~pc, respectively. Our convergent point method yields a Columba membership likelihood of 92\%, while BANYAN yields 69\% (86\% when including the radial velocity measurement). 
We calculate UVW velocities adopting a distance of $\sim$100~pc for this system assuming Columba membership (see Table~\ref{tab:uvw}). 
The UVW velocity uncertainties are high given the distance uncertainty of 20\%, but the UVW are a good match to those listed in \citet{Torres:2008} for Columba.
Given that spectroscopic signatures point to a young age ($\gtrsim$10~Myr), J0259--42 may be a new member of the $\sim$30~Myr-old Columba Association, in agreement with kinematic indicators.

{\it J0315--53:} As mentioned in Section~\ref{spec1res}, this M5 system has strong Li absorption, indicative of a young age. 
However, the kinematics of the system agree with those of Tuc-Hor, which in turn appears inconsistent with the Li measurement. 
We have measured a radial velocity of $9.4\pm1.5$~km/s with an R$\sim$7000 WiFeS spectrum. Table~\ref{tab:uvw} lists calculated UVW velocities for the system assuming a kinematic distance of 49~pc. As Figure~\ref{fig:uvw} illustrates, the UVW velocities match those of known Tuc-Hor members within the uncertainties.
If J0315--53 were $\sim$10~Myr-old, our empirical isochrones suggest a distance of 65~pc. However, the UVW velocities J0315--53 would have at that distance ($-10.7\pm3.2$, $-23.7\pm4.3$, $6.6\pm3.5$~km/s) do not match those of any known young groups.
We note that this is not the first time a Li-rich low-mass star has been classified as a potential member of Tuc-Hor (see the discussion of J0236--52 above) and, hence, J0315--53 may indeed be a new Tuc-Hor member. 
A parallax distance measurement will be needed to fully confirm membership of this low-mass star.

{\it J0324--27:} Like J0259--42, J0324--27 also displays a strong infrared excess (Figure~\ref{fig:sed}); its 
$W1-W4$ color is 4.2. The spectral type based on the J--W2 color (M9) is overestimated due to this excess, as 
is the case for J0259--42. Near-IR spectroscopy (Figure~10) suggests a spectral type intermediate 
between M5 and M6; we adopt M5.5. Furthermore, the SpeX spectra show some signatures of youth, such as 
Br-$\gamma$ emission and weaker Na~I absorption than the field, but, like J0259--42, J0324--27 lacks the 
low-surface-gravity H-band profile. 
Warm dust grains appear to be present in the system and also point towards a youthful nature. 
The kinematic distance estimate ($\sim$110~pc) puts the star too far away for membership in Tuc-Hor. 
Columba membership, on the other hand, is likely: 60-80\% likelihood, with a distance of 90--110~pc.
Overall, the system is very similar to J0259--42 and may be a young Columba member.

{\it J0413--44:} This M4 system was observed as part of the RAVE survey and has a published radial velocity of 
$2.3\pm6.6$~km/s \citep{Siebert:2011}. A radial velocity closer to $\sim$16~km/s would be required for 
Tuc-Hor membership as seen in Table~\ref{table:th2}. The measured radial velocity, combined with our estimated kinematic distance (56~pc), allows us to estimate UVW velocities of 
$-5.9\pm2.0$, $-11.0\pm4.7$, and $8.6\pm5.3$~km/s (see Table~\ref{tab:uvw}).
Hence, the UVW velocities for J0413--44 differ significantly from those of Tuc-Hor. 
With such a large, positive W velocity, it stands apart from many 
of the known young moving groups (see Figure~\ref{fig:uvw}). 
Indeed, given its measured radial velocity, we can find no distance for J0413--44 that would place it in the good UVW box.  
Therefore, while initially labeled as a Tuc-Hor candidate given its proper motions, J0413--44 is not a member of Tuc-Hor and, furthermore, might not be a young star.

% 60% Old
{\it J0105--48, J0205--60, J0213--46, J0217--30, J0217--32:} BANYAN predicts these systems lie among the field (old) population, rather than belonging to any particularly young moving group. 
Our convergent point analysis predicts a high likelihood of AB~Dor membership (93\%) for J0217--30 and J0217--32, and a high likelihood of $\beta$~Pic membership (87\%) for J0213--46. However, the kinematic distances for these three stars (89, 97, 76~pc) are on the high end of what is typically observed for members of these groups and also differ when compared to the 10 and 100~Myr isochrone distances.
Spectroscopy for these candidates will be necessary to search for features of youth and confirm whether or not they are old.

% 60% Columba
{\it J0142--51, J0233--18, J0305--37, J0308-38, J0321--33, J0339--24, J0407--29, J0427--24, J0427--33, J0431--30:} 
These systems are similar to J0202--31, J0259--42, and J0324--27 in the sense of high membership likelihoods ($>$60\%) for the Columba Association according to BANYAN. In many cases, the convergent point also returns a high likelihood of membership in Columba.
Two systems, however, have low convergent point likelihoods of Columba membership (J0154--29:~21\% and J0341--22:~45\%). To be conservative, we do not list these two systems in Table~\ref{tab:final}.
The other systems listed here remain viable candidates, but accurate radial velocities and distances will be necessary to fully vet their membership.

% ~40% in a pair of groups: 
{\it J0221--58, J0336--26:} These two systems have equal BANYAN likelihoods ($\sim$40\%) for membership in two separate groups. 
J0221--58 may belong to either AB~Dor or the old field population. The convergent point predicts a high probability of membership for AB~Dor ($\sim$95\%). This M3.5 dwarf shows X-ray emission, but further work will be needed to ascertain the membership of the system.
J0336--26, on the other hand, may belong to either Tuc-Hor or Columba, according to BANYAN. Our convergent point predicts equally high ($\sim$70\%) likelihood of membership in Tuc-Hor and Columba. A kinematic distance of $\sim$50~pc is predicted for either group, which is the same as what we estimate with the 10-Myr isochrone.
Given the likelihoods of belonging to two separate groups, we hold back on placing either system in any young moving group until additional information is obtained.

% Low CP, but high/good BANYAN
{\it J0316--35, J0320--50:} These systems have very low convergent point likelihoods ($<$10\%) of belonging to Tuc-Hor. However, BANYAN predicts moderate to high likelihood of membership (70--90\%). 
J0236--52, discussed above, is another such system where the convergent point analysis returns a low membership likelihood. 
Given the low convergent point likelihoods, we do not list them among plausible Tuc-Hor candidates in Table~\ref{tab:final}.

% 50% beta Pic
{\it J0318--34, J0352--28:} These objects have low likelihoods of Tuc-Hor membership, and higher for $\beta$~Pic in both BANYAN and the convergent point methods.
However, the 10-Myr isochrone distances for these stars (80, 100~pc) are much larger than the $\beta$~Pic kinematic distances we estimate (50, 30~pc). 
Spectroscopy will be required to examine whether these stars are young enough to be $\beta$~Pic members as implied by their kinematics. However, given the large distance discrepancies, we anticipate that these stars are unlikely to be new $\beta$~Pic members.

\clearpage

% Selection Criteria
\begin{table}
\begin{center}
\begin{tabular}{ll}
\hline
\multicolumn{2}{c}{Selection Criteria}  \\
\hline
(1) & $9.5 \leq NUV-W1 < 12.5$ \\
(2) & $J-W2 \geq 0.8 $ \\
(3) & $0 < W1-W2 < 0.6 $ \\
(4) & $\text{n\_2mass} = 1$ \\
(5) & $W2\geq6\; $ \\
(6) & $W2\leq14$  \\
(7) & $W2\leq12 \text{ for } 1.7<J-W2<3.3 $ \\
\hline\end{tabular}
\caption{Selection criteria used for our GALEX-WISE-2MASS search. 
The first 3 expressions are used to select UV-bright dwarfs with spectral types late-K to early-L. 
The remaining criteria facilitate our search by filtering out crowded regions, saturated stars, and possible faint galaxies (see details in Section~\ref{method:uv}).
 } \label{select_criteria}
 \end{center}
\end{table}

\clearpage

% TH Cands
\begin{table}
\begin{center}
\begin{tabular}{rccccccl}
%\begin{longtable}{rccccccl}

%\multicolumn{8}{c}
%{\tablename\ \thetable\ -- Candidate Young Stars} \\

\hline
Index 	& WISE 		& $\mu_{RA}$ 	& $\mu_{Dec}$ 	& $\mu_{tot}$ 	& PM 	& Est. Sp. 	& F-Value \\
		& Designation	& (mas/yr)	& (mas/yr)		& (mas/yr)	& Source	& Type$^a$	& \\
\hline

1	&	J010516.16-484116.9	& $46.8\pm1.7$	& $-15.6\pm1.7$ &	49.3	&	UCAC4	& M5.5 &	1.55	\\
2	&	J012758.87-603224.5	& $89.9\pm3.1$	& $-30.4\pm2.9$ &	94.9	&	UCAC4	& M4.1 &	1.04	\\
3	&	J014246.89-512646.9	& $66.8\pm4.2$	& $-12.7\pm4.2$ &	68.0	&	UCAC4	& M6.7 &	0.68	\\
4	&	J015057.01-584403.4	& $92.2\pm2.0$	& $-24.3\pm2.0$ &	95.3	&	UCAC4	& M2.8 &	1.59	\\
5	&	J015325.09-683322.8	& $98.0\pm2.9$	& $-15.1\pm2.4$ &	99.2	&	UCAC4	& M5.2 &	0.36	\\
6	&	J015455.24-295746.0	& $78.7\pm2.0$	& $-23.6\pm1.4$ &	82.2	&	UCAC4	& M5.2 &	0.79	\\
7	&	J020020.08-661402.0	& $84.0\pm6.3$	& $-11.4\pm3.4$ &	84.8	&	UCAC4	& M4.4 &	0.68	\\
8	&	J020257.94-313626.4	& $81.0\pm2.9$	& $-27.7\pm2.3$ &	85.6	&	UCAC4	& M4.7 &	0.47	\\
9$^b$	&	J020306.68-554542.1	& $54.2\pm25.0$	& $-10.7\pm25.0$ &	55.2	&	WISE-2M	& M4.5 &	0.48	\\
10	&	J020547.70-602808.4	& $89.1\pm1.5$	& $-61.7\pm1.5$ &	108.4	&	UCAC4	& M5.0 &	0.39	\\
11	&	J020701.85-440638.3	& $94.9\pm1.3$	& $-30.6\pm1.3$ &	99.7	&	UCAC4	& M3.3 &	1.57	\\
12$^c$	&	J021053.50-460351.4	& $53.2\pm1.8$	& $-10.2\pm1.8$ &	54.2	&	UCAC4	& M4.3 &	0.82	\\
13	&	J021258.28-585118.3	& $87.7\pm1.3$	& $-15.9\pm1.3$ &	89.1	&	UCAC4	& M2.8 &	0.38	\\
14	&	J021330.24-465450.3	& $42.5\pm1.0$	& $4.9\pm1.0$ &	42.8	&	UCAC4	& M3.8 &	1.07	\\
15	&	J021533.37-562717.6	& $86.4\pm17.1$	& $-24.7\pm17.1$ &	89.8	&	PPMXL	& M7.5 &	0.83	\\
16	&	J021705.03-300621.9	& $40.7\pm4.5$	& $-36.1\pm4.0$ &	54.4	&	UCAC4	& M4.6 &	1.26	\\
17	&	J021745.82-321718.2	& $37.4\pm13.4$	& $-31.0\pm0.9$ &	48.6	&	UCAC4	& M0.7 &	0.56	\\
18	&	J022051.50-582341.3	& $97.3\pm2.0$	& $-13.0\pm2.0$ &	98.2	&	UCAC4	& M3.2 &	0.62	\\
19	&	J022142.84-583204.4	& $46.4\pm2.4$	& $-2.2\pm2.4$ &	46.5	&	UCAC4	& M3.5 &	0.83	\\
20	&	J022244.32-602247.7	& $137.4\pm1.7$	& $-13.8\pm1.7$ &	138.1	&	UCAC4	& M3.8 &	1.04	\\
21	&	J022424.69-703321.2	& $92.5\pm2.7$	& $-3.6\pm3.9$ &	92.6	&	UCAC4	& M4.3 &	0.35	\\
22	&	J023219.44-574611.9	& $83.8\pm2.3$	& $-17.1\pm2.6$ &	85.5	&	UCAC4	& M4.7 &	0.29	\\
23	&	J023359.89-181152.5	& $53.5\pm1.7$	& $-22.5\pm1.7$ &	58.0	&	UCAC4	& M3.7 &	1.24	\\
24	&	J023651.80-520303.5	& $102.2\pm0.8$	& $1.2\pm0.8$ &	102.2	&	UCAC4	& M2.6 &	0.85	\\
25	&	J024127.29-304915.1	& $97.0\pm2.1$	& $-28.0\pm2.2$ &	101.0	&	UCAC4	& M4.6 &	0.91	\\
26	&	J024202.14-535914.7	& $97.0\pm2.1$	& $-20.2\pm2.2$ &	99.1	&	UCAC4	& M4.5 &	0.20	\\
27	&	J024204.15-535900.0	& $98.4\pm2.2$	& $-9.1\pm7.6$ &	98.8	&	UCAC4	& M4.8 &	0.47	\\
28	&	J024746.49-580427.4	& $95.5\pm1.4$	& $-5.2\pm3.5$ &	95.6	&	UCAC4	& M2.9 &	0.94	\\
29	&	J025022.35-654555.2	& $75.8\pm1.9$	& $3.4\pm1.9$ &	75.9	&	UCAC4	& M3.4 &	1.26	\\
30	&	J025059.67-340905.3	& $87.2\pm1.8$	& $-21.0\pm1.8$ &	89.7	&	UCAC4	& M4.6 &	1.06	\\
31	&	J025433.25-510831.4	& $92.0\pm1.2$	& $-11.9\pm1.2$ &	92.8	&	UCAC4	& M3.1 &	0.47	\\
32	&	J025531.87-570252.3	& $90.5\pm2.7$	& $-8.0\pm2.8$ &	90.9	&	UCAC4	& M4.2 &	0.70	\\
33	&	J025901.49-423220.4	& $39.9\pm4.1$	& $-6.9\pm4.4$ &	40.5	&	UCAC4	& M7.4 &	0.55	\\
34	&	J030505.65-531718.4	& $91.4\pm3.6$	& $-11.0\pm3.6$ &	92.1	&	UCAC4	& M4.8 &	1.04	\\
35	&	J030509.79-372505.8	& $50.8\pm1.3$	& $-12.2\pm1.3$ &	52.2	&	UCAC4	& M2.5 &	0.66	\\
36	&	J030839.55-384436.3	& $68.7\pm3.2$	& $-10.1\pm4.2$ &	69.4	&	UCAC4	& M4.2 &	0.64	\\
37	&	J031049.48-361647.3	& $90.4\pm1.9$	& $-28.1\pm1.9$ &	94.7	&	UCAC4	& M4.3 &	0.37	\\
38	&	J031145.52-471950.2	& $88.9\pm1.8$	& $-3.6\pm2.0$ &	89.0	&	UCAC4	& M4.5 &	0.68	\\
39	&	J031523.72-534253.9	& $81.0\pm9.5$	& $-10.8\pm7.2$ &	81.7	&	UCAC4	& M5.2 &	1.20	\\
40	&	J031650.45-350937.9	& $92.3\pm1.1$	& $-38.3\pm1.1$ &	99.9	&	UCAC4	& M3.7 &	0.79	\\
41	&	J031856.73-343317.6	& $44.5\pm2.6$	& $7.9\pm2.8$ &	45.2	&	UCAC4	& M4.3 &	0.96	\\
42	&	J032047.66-504133.0	& $82.6\pm1.7$	& $7.8\pm1.5$ &	83.0	&	UCAC4	& M2.2 &	1.68	\\
43	&	J032144.76-330949.5	& $40.5\pm3.1$	& $-13.6\pm3.1$ &	42.7	&	UCAC4	& M6.0 &	1.16	\\
44	&	J032440.63-390422.8	& $86.3\pm1.9$	& $-17.4\pm1.6$ &	88.0	&	UCAC4	& M4.2 &	0.56	\\
45	&	J032443.06-273323.1	& $34.4\pm3.9$	& $-13.6\pm3.9$ &	36.9	&	PPMXL	& M9.2 &	0.54	\\
46	&	J032916.57-370250.2	& $82.2\pm2.6$	& $-21.6\pm2.2$ &	85.0	&	UCAC4	& M4.4 &	0.31	\\
47	&	J033631.50-261958.1	& $81.0\pm4.0$	& $-19.0\pm4.0$ &	83.2	&	UCAC4	& M5.7 &	0.98	\\
48	&	J033901.64-243406.1	& $66.7\pm2.6$	& $-17.6\pm2.7$ &	69.0	&	UCAC4	& M6.0 &	0.41	\\
49	&	J034115.60-225307.8	& $51.9\pm2.3$	& $-14.2\pm1.6$ &	53.8	&	UCAC4	& M1.4 &	0.74	\\
50	&	J035122.95-515458.1	& $71.7\pm1.8$	& $4.2\pm1.8$ &	71.8	&	UCAC4	& M4.2 &	0.46	\\
51	&	J035223.52-282619.6	& $70.5\pm1.0$	& $-1.7\pm1.0$ &	70.5	&	UCAC4	& M2.0 &	1.01	\\
52	&	J035616.31-391521.8	& $67.7\pm2.2$	& $-4.9\pm2.2$ &	67.9	&	UCAC4	& M4.2 &	0.06	\\
53	&	J040539.68-401410.5	& $71.6\pm2.0$	& $-0.8\pm2.1$ &	71.6	&	UCAC4	& M4.2 &	0.89	\\
54	&	J040711.50-291834.3	& $42.0\pm1.1$	& $-6.9\pm1.0$ &	42.6	&	UCAC4	& M1.0 &	0.84	\\
55	&	J041336.14-441332.4	& $56.2\pm1.7$	& $0.7\pm2.1$ &	56.2	&	UCAC4	& M3.9 &	0.59	\\
56	&	J042726.28-245527.4	& $54.8\pm3.9$	& $-14.6\pm3.9$ &	56.7	&	PPMXL	& M4.5 &	0.50	\\
57	&	J042745.66-332742.6	& $48.4\pm3.3$	& $-1.0\pm4.7$ &	48.4	&	UCAC4	& M4.5 &	1.00	\\
58	&	J043138.61-304250.9	& $33.7\pm1.4$	& $-2.2\pm1.4$ &	33.8	&	UCAC4	& M3.5 &	0.80	\\
\hline
\end{tabular}
\end{center}
\caption{
Candidate young, low-mass stars selected with the F-value analysis (see Section~\ref{fvalue}) in the vicinity of Tuc-Hor. 
\\ Notes: $^a$ Spectral types estimated from J--W2 color (see Section~\ref{spectypes}).
\newline $^b$ PPMXL and UCAC4 proper motions for J0203-55 differ substantially; we use WISE-2MASS estimated proper motions.
\newline $^c$ J0210--46 is a low-mass companion to AB Dor member CD-46~644 (see Appendix~\ref{notesys}).
} \label{table:th1}
\end{table}
%\end{longtable}

\clearpage

% Convergent Point information
\begin{table}
\begin{center}
\begin{tabular}{lccccc}
\hline
Moving  &
  \multicolumn{2}{c}{Conv.\ Point} &
Velocity &
Group &
$\sigma_{int}$ \\
Distance &
RA & Dec &
(km/s) &
\ (pc) &
(km s$^{-1}$) \\
\hline
Tuc-Hor & 119 & --27 & 23.3 & 50 & 1 \\
$\beta$~Pic & 90 & --28 & 20.8 & 40 & 1 \\
AB~Dor & 92 & --47 & 31.2 & 50 & 2 \\
TWA & 95 & --26 & 21.6 & 50 & 1 \\
Carina-Near & 98 & 0 & 31.3 & 30 & 3\\ %2.6
Columba & 106 & --30 & 26.5 & 80 & 1\\
\hline\end{tabular}
\caption{Convergent point estimates for several young moving groups within 100~pc of the Earth. 
The listed velocities are the total space motion with respect to the Sun.
These values can be used to estimate membership probabilities, kinematic distances, and radial velocities as described in Section~\ref{cpanalysis}. 
We adopt a 2~degree uncertainty on the convergent point location for all groups (Section~\ref{cpanalysis}).
As described in \citet{Mamajek:2005}, the information returned from the convergent point analysis is not 
particularly sensitive to group distance and internal dispersion ($\sigma_{int}$). We note that the TWA 
convergent point listed here is somewhat different than that used in \citet{Looper:2010b} (99.8, $-27.7$ 
degrees with 22.0 km/s full space velocity).
Carina-Near members are drawn from \citet{Zuckerman:2006}; the others are from \citet{Torres:2008}, \citet{Zuckerman:2011}, and \citet{Schneider:2012a}.
 } \label{cpinfo}
\end{center}
\end{table}

\clearpage

% TucHor Membership likelihood (tab2_long)

\begin{table}
\begin{center}
\begin{tabular}{l|rrr|rrr}
%\begin{longtable}{l|rrr|rrr}

\hline
  \multicolumn{1}{c|}{} & \multicolumn{3}{c|}{This work} & \multicolumn{3}{c}{BANYAN$^a$} \\
\hline
  \multicolumn{1}{c|}{ID} & \multicolumn{1}{c}{Prob} & \multicolumn{1}{c}{RV} & \multicolumn{1}{c|}{Dist} & \multicolumn{1}{c}{Prob} & \multicolumn{1}{c}{RV} & \multicolumn{1}{c}{Dist} \\
%\hline
  \multicolumn{1}{c|}{} & \multicolumn{1}{c}{(\%)} & \multicolumn{1}{c}{(km/s)} & \multicolumn{1}{c|}{(pc)} & \multicolumn{1}{c}{(\%)} & \multicolumn{1}{c}{(km/s)} & \multicolumn{1}{c}{(pc)} \\
\hline

  J0105--48$^b$ & 30.2 & 4.9 & 98.1 & 0.0 & 6.0 & \nodata \\
  J0127--60 & 95.8 & 7.9 & 48.5 & 97.9 & 8.9 & 46.5\\
  J0142--51$^b$ & 23.6 & 7.5 & 69.0 & 16.9 & 8.5 & 55.0\\
  J0150--58 & 95.4 & 8.8 & 47.5 & 96.8 & 9.7 & 45.5\\
  J0153--68 & 79.0 & 9.7 & 44.9 & 99.0 & 10.6 & 44.0\\
  J0154--29$^b$ & 15.0 & 5.2 & 58.5 & 8.4 & 6.3 & 49.5\\
  J0200--66 & 78.8 & 9.8 & 52.4 & 98.4 & 10.7 & 49.5\\
  J0202--31$^b$ & 60.7 & 6.1 & 55.3 & 20.5 & 7.1 & 48.5\\
  J0203--55$^b$ & 99.1 & 9.1 & 81.7 & 8.3 & 10.0 & 60.0\\
  J0205--60$^b$ & 0.0 & 9.6 & 44.5 & 0.5 & 10.5 & \nodata \\ 
  J0207--44 & 98.7 & 8.0 & 46.0 & 98.5 & 9.1 & 43.5\\
  J0210--46$^b$ & 49.5 & 8.5 & 84.5 & 0.2 & 9.6 & \nodata \\
  J0212--58 & 95.3 & 9.8 & 49.8 & 96.3 & 10.8 & 47.0\\
  J0213--46$^b$ & 0.1 & 8.8 & 115.0 & 0.0 & 9.8 & \nodata \\
  J0215--56 & 94.0 & 9.8 & 49.6 & 94.9 & 10.7 & 46.5\\
  J0217--30$^b$ & 1.0 & 6.9 & 91.3 & 0.0 & 8.0 & \nodata \\
  J0217--32$^b$ & 6.2 & 7.3 & 101.0 & 0.0 & 8.3 & \nodata \\
  J0220--58 & 78.3 & 10.1 & 44.9 & 95.7 & 11.1 & 43.5\\
  J0221--58$^b$ & 54.0 & 10.2 & 95.4 & 0.4 & 11.1 & \nodata \\
  J0222--60 & 57.6 & 10.3 & 31.8 & 65.1 & 11.3 & 32.5\\
  J0224--70 & 98.0 & 10.8 & 46.8 & 99.6 & 11.6 & 46.0\\
  J0232--57 & 65.5 & 10.6 & 50.9 & 95.2 & 11.6 & 47.5\\
  J0233--18$^b$ & 98.5 & 6.5 & 80.9 & 0.0 & 7.5 & \nodata \\
  J0236--52$^b$ & 0.3 & 10.6 & 43.4 & 60.6 & 11.5 & 41.0\\
  J0241--30 & 69.8 & 8.9 & 44.8 & 59.4 & 9.9 & 42.5\\
  J0242--53W & 54.7 & 10.9 & 43.6 & 92.0 & 11.9 & 42.0\\
  J0242--53E & 87.9 & 10.9 & 43.7 & 96.5 & 11.9 & 42.5\\
  J0247--58 & 92.3 & 11.4 & 44.6 & 95.0 & 12.3 & 43.5\\
  J0250--65 & 85.4 & 11.6 & 55.9 & 97.5 & 12.4 & 51.5\\
  J0250--34 & 70.2 & 10.0 & 49.3 & 58.1 & 10.9 & 45.5\\
  J0254--51 & 99.9 & 11.5 & 45.8 & 97.0 & 12.4 & 44.0\\
  J0255--57 & 93.3 & 11.7 & 46.5 & 94.6 & 12.6 & 44.5\\
  J0259--42$^b$ & 98.6 & 11.3 & 105.7 & 0.0 & 12.2 & \nodata \\
  J0305--53 & 77.1 & 12.1 & 45.4 & 92.9 & 13.0 & 43.5\\
  J0305--37$^b$ & 99.8 & 11.3 & 81.9 & 0.6 & 12.2 & \nodata \\
  J0308--38$^b$ & 78.7 & 11.6 & 61.1 & 23.8 & 12.5 & 52.0\\
  J0310--36 & 48.6 & 11.6 & 44.9 & 95.9 & 12.5 & 43.0\\
  J0311--47 & 52.2 & 12.3 & 46.8 & 91.8 & 13.2 & 44.0\\
  J0315--53 & 63.5 & 12.6 & 50.5 & 86.6 & 13.5 & 46.5\\
  J0316--35$^b$ & 1.2 & 11.9 & 42.6 & 93.4 & 12.8 & 40.5\\
  J0318--34$^b$ & 0.3 & 12.0 & 100.8 & 0.0 & 12.9 & \nodata \\
  J0320--50$^b$ & 9.3 & 12.9 & 49.5 & 71.8 & 13.7 & 45.5\\
  J0321--33$^b$ & 80.0 & 12.1 & 98.1 & 0.0 & 13.0 & \nodata \\
  J0324--39 & 88.8 & 12.7 & 46.6 & 94.4 & 13.6 & 44.0\\
  J0324--27$^b$ & 86.6 & 11.8 & 114.5 & 0.0 & 12.6 & \nodata \\
  J0329--37 & 51.7 & 12.9 & 48.0 & 95.0 & 13.7 & 45.0\\
  J0336--26$^b$ & 76.2 & 12.5 & 49.6 & 44.1 & 13.3 & 46.5\\
  J0339--24$^b$ & 84.8 & 12.5 & 59.7 & 19.1 & 13.3 & 52.5\\
  J0341--22$^b$ & 84.6 & 12.5 & 76.6 & 1.2 & 13.3 & 59.5\\
  J0351--51 & 97.6 & 14.4 & 53.3 & 93.3 & 15.2 & 49.0\\
  J0352--28$^b$ & 0.9 & 13.9 & 57.0 & 20.0 & 14.6 & 49.5\\
  J0356--39 & 97.7 & 14.7 & 55.8 & 65.4 & 15.4 & 50.0\\
  J0405--40 & 87.0 & 15.3 & 51.5 & 68.8 & 16.0 & 47.5\\
  J0407--29$^b$ & 90.9 & 14.9 & 88.1 & 0.2 & 15.6 & \nodata \\
  J0413--44 & 96.9 & 15.7 & 64.1 & 81.4 & 16.4 & 54.0\\
  J0427--24$^b$ & 99.6 & 15.9 & 62.7 & 14.8 & 16.5 & 55.5\\
  J0427--33$^b$ & 80.8 & 16.4 & 71.7 & 3.6 & 17.0 & 57.5\\
  J0431--30$^b$ & 82.6 & 16.5 & 102.0 & 0.0 & 17.1 & \nodata \\
\hline
\end{tabular}
\end{center}
\caption{Convergent point analysis probabilities, predicted radial velocities, and kinematic distances for our Tuc--Hor candidates (see Table~\ref{table:th2}). BANYAN does not predict distances for low probability objects. 
\\ Notes: $^a$ See \citet{Malo:2012}.
\newline $^b$ Possible members of other moving groups or old field dwarfs, see Appendix~\ref{notesys}.
} \label{table:th2}
\end{table}

\clearpage

% Spectroscopic measurements
\begin{table}
\begin{center}
\begin{tabular}{llrrrr}
\hline
Object & Sp.\ Type & H$\alpha$ EW (\AA) & Li EW (\AA) & Na I EW (\AA) & Na I Index \\
\hline
J0202--31	&	M4.0	&	$-4.34\pm0.26$	&	$<0.15$	&	$4.41\pm0.32$	&	1.22	\\
J0212--58	&	M2.1	&	$-4.67\pm0.27$	&	$<0.01$	&	$2.92\pm0.29$	&	1.13	\\
J0215--56	&	M5.4	&	$-10.29\pm0.78$	&	$<0.03$	&	$3.99\pm0.45$	&	1.20	\\
J0220--58	&	M3.0	&	$-7.38\pm0.72$	&	$<0.06$	&	\nodata	&	\nodata	\\
J0232--57	&	M4.4	&	$-6.07\pm0.34$	&	$<0.05$	&	$3.65\pm0.30$	&	1.16	\\
J0241--30	&	M4.7	&	$-8.86\pm0.35$	&	$<0.06$	&	$3.98\pm0.36$	&	1.21	\\
J0242--53W	&	M4.6	&	$-9.65\pm0.28$	&	$<0.05$	&	$3.77\pm0.31$	&	1.19	\\
J0250--65	&	M3.7	&	$-7.19\pm0.34$	&	$<0.02$	&	$3.24\pm0.29$	&	1.16	\\
J0255--57	&	M4.9	&	$-7.92\pm0.33$	&	$<0.02$	&	$4.14\pm0.35$	&	1.20	\\
J0259--42	&	M4.2	&	$-11.01\pm0.89$	&	$<0.11$	&	$3.56\pm0.48$	&	1.17	\\ 
J0305--37	&	M1.9	&	$-4.18\pm0.25$	&	$<0.01$	&	$2.55\pm0.23$	&	1.12	\\
J0305--53	&	M5.4	&	$-10.25\pm0.43$	&	$<0.05$	&	$4.17\pm0.33$	&	1.19	\\
J0311--47	&	M4.3	&	$-4.51\pm0.27$	&	$<0.04$	&	$3.35\pm0.35$	&	1.17	\\
J0315--53	&	M5.2	&	$-7.53\pm0.52$	&	$0.37\pm0.05$	&	$3.99\pm0.37$	&	1.21	\\
J0318--34	&	M4.1	&	$-5.63\pm0.32$	&	$<0.02$	&	$4.38\pm0.31$	&	1.24	\\
J0320--50	&	M2.0	&	$-1.28\pm0.21$	&	$<0.03$	&	$2.15\pm0.31$	&	1.11	\\
J0329--37	&	M4.3	&	$-8.63\pm0.34$	&	$<0.03$	&	$3.35\pm0.36$	&	1.17	\\
J0351--51	&	M4.4	&	$-7.70\pm0.44$	&	$<0.03$	&	$3.71\pm0.32$	&	1.19	\\
J0356--39	&	M5.0	&	$-10.01\pm0.41$	&	$<0.04$	&	$3.86\pm0.32$	&	1.19	\\
J0405--40	&	M4.2	&	$-8.40\pm0.36$	&	$<0.02$	&	$3.44\pm0.31$	&	1.17	\\
J0413--44	&	M3.9	&	$-9.47\pm0.35$	&	$<0.03$	&	$3.37\pm0.31$	&	1.17	\\
J0427--33	&	M4.8	&	$-9.63\pm0.36$	&	$<0.04$	&	$4.13\pm0.32$	&	1.21	\\
J0431--30	&	M3.2	&	$-7.51\pm0.31$	&	$<0.03$	&	$3.07\pm0.36$	&	1.15	\\
\hline
J0207--44$^a$&	M3.5	&	$-4.1$	&	\nodata	&	\nodata	&	\nodata	\\
J0213--46$^a$&	M4	&	$-8.6$	&	\nodata	&	\nodata	&	\nodata	\\
J0222--60$^a$&	M4	&	$-8.1$	&	\nodata	&	\nodata	&	\nodata	\\
J0233--18$^a$&	M3	&	$-8.5$	&	\nodata	&	\nodata	&	\nodata	\\
J0236--52$^b$&	M2	&	$-5.3$	&	$0.32$	&	\nodata	&	\nodata	\\
J0254--51$^a$&	M1.5	&	$-3.1$	&	\nodata	&	\nodata	&	\nodata	\\
J0407--29$^a$&	M0	&	$-3.2$	&	\nodata	&	\nodata	&	\nodata	\\
\hline\end{tabular}
\caption{WiFeS spectroscopic measurements for some Table~\ref{table:th1} Tuc-Hor candidates. 
We measure equivalent widths (EW) for H$\alpha$, Li, and the sum of the two lines in the Na~I doublet at 8183 and 8195\AA. 
Spectra for J0220--58 were taken with R$\sim$7000 and do not cover the Na~I region.
Spectral types are determined using the TiO5 index \citet{Reid:1995} and are accurate to $\pm$0.5. 
The Na~I index is the ratio of the average flux on and off the doublet (Figure~\ref{naspec}) and is measured in spectra that have been re-binned to R$\sim$800 (see Section~\ref{sodium}). 
The final 7 systems have measurements published in the literature 
($^a$: \citealt{Riaz:2006}, $^b$: \citealt{Torres:2006}).} \label{meas1}
\end{center}
\end{table}

\clearpage

% X-ray data
\begin{table}
\begin{center}
\begin{tabular}{lrccrcc}
%\begin{longtable}{lrccrcc}

\hline
Object & Dist.\ & log $L_{bol}/L_\odot$ & X-ray Counterpart & Offset & X-ray Count & log $L_X/L_{bol}$ \\
& (pc) &  & 1RXS & (\arcsec) & Rate (s$^{-1}$) & \\
\hline

\hline
J0207--44	&	46	&	-1.17	&	 J020701.9-440645	&	7	&	$0.076\pm0.014$	&	-3.23	\\
J0212--58	&	50	&	-1.09	&	 J021257.7-585109	&	10	&	$0.185\pm0.053$	&	-2.84	\\
J0213--46	&	115	&	-0.44	&	 J021329.9-465452	&	4	&	$0.106\pm0.017$	&	-3.01	\\
J0220--58	&	45	&	-1.53	&	 J022052.8-582328	&	16	&	$0.056\pm0.020$	&	-3.01	\\
J0221--58	&	95	&	-1.03	&	 J022145.6-583159	&	22	&	$0.034\pm0.016$	&	-3.08	\\
J0222--60	&	32	&	-1.36	&	 J022243.9-602243	&	6	&	$0.357\pm0.042$	&	-2.68	\\
J0233--18	&	81	&	-1.00	&	 J023400.1-181155	&	4	&	$0.102\pm0.020$	&	-2.78	\\
J0236--52	&	43	&	-0.86	&	 J023651.8-520300	&	4	&	$0.332\pm0.028$	&	-2.95	\\
J0241--30	&	45	&	-1.92	&	 J024127.5-304921	&	7	&	$0.029\pm0.010$	&	-2.91	\\
J0242--53E	&	44	&	-1.60	&	 J024202.5-535908	&	17	&	$0.036\pm0.012$	&	-3.17	\\
J0242--53W	&	44	&	-1.87	&	 J024202.5-535908	&	7	&	$0.036\pm0.012$	&	-2.89	\\
J0250--34	&	49	&	-1.60	&	 J025100.3-340914	&	12	&	$0.066\pm0.016$	&	-2.81	\\
J0254--51	&	46	&	-0.90	&	 J025432.4-510829	&	8	&	$0.122\pm0.022$	&	-3.28	\\
J0305--37	&	82	&	-0.76	&	 J030510.1-372507	&	4	&	$0.050\pm0.015$	&	-3.32	\\
J0308--38	&	61	&	-1.74	&	 J030840.1-384439	&	7	&	$0.059\pm0.014$	&	-2.53	\\
J0311--47	&	47	&	-1.62	&	 J031145.2-471936	&	14	&	$0.028\pm0.010$	&	-3.19	\\
J0324--39	&	47	&	-1.40	&	 J032439.7-390421	&	11	&	$0.395\pm0.039$	&	-2.27	\\
J0352--28	&	57	&	-1.19	&	 J035222.3-282610	&	19	&	$0.070\pm0.014$	&	-3.05	\\
J0407--29	&	88	&	-0.48	&	 J040710.6-291823	&	16	&	$0.137\pm0.019$	&	-3.10	\\
J0431--30	&	102	&	-0.82	&	 J043137.9-304237	&	16	&	$0.059\pm0.018$	&	-2.99	\\
\hline													
J0105--48	&	98	&	-1.66	&		&		&		&	$<$-2.56	\\
J0127--60	&	48	&	-1.86	&		&		&		&	$<$-2.98	\\
J0142--51	&	69	&	-1.53	&		&		&		&	$<$-2.99	\\
J0150--58	&	48	&	-1.22	&		&		&		&	$<$-3.62	\\
J0153--68	&	45	&	-1.91	&		&		&		&	$<$-3.00	\\
J0154--29	&	58	&	-2.23	&		&		&		&	$<$-2.45	\\
J0200--66	&	52	&	-1.65	&		&		&		&	$<$-3.13	\\
J0202--31	&	55	&	-1.95	&		&		&		&	$<$-2.77	\\
J0203--55	&	82	&	-1.47	&		&		&		&	$<$-2.91	\\
J0205--60	&	44	&	-1.67	&		&		&		&	$<$-3.24	\\
J0210--46	&	84	&	-1.47	&		&		&		&	$<$-2.88	\\
J0215--56	&	50	&	-2.14	&		&		&		&	$<$-2.67	\\
J0217--30	&	91	&	-1.13	&		&		&		&	$<$-3.16	\\
J0217--32	&	101	&	-0.25	&		&		&		&	$<$-3.95	\\
J0224--70	&	47	&	-1.57	&		&		&		&	$<$-3.28	\\
J0232--57	&	51	&	-1.81	&		&		&		&	$<$-2.98	\\
J0247--58	&	45	&	-1.19	&		&		&		&	$<$-3.71	\\
J0250--65	&	56	&	-1.40	&		&		&		&	$<$-3.30	\\
J0255--57	&	46	&	-1.96	&		&		&		&	$<$-2.92	\\
J0259--42	&	106	&	-1.63	&		&		&		&	$<$-2.53	\\
J0305--53	&	45	&	-1.97	&		&		&		&	$<$-2.92	\\
J0310--36	&	45	&	-1.76	&		&		&		&	$<$-3.15	\\
J0315--53	&	51	&	-1.90	&		&		&		&	$<$-2.90	\\
J0316--35	&	43	&	-1.20	&		&		&		&	$<$-3.74	\\
J0318--34	&	101	&	-1.13	&		&		&		&	$<$-3.07	\\
J0320--50	&	50	&	-1.15	&		&		&		&	$<$-3.65	\\
J0321--33	&	98	&	-1.30	&		&		&		&	$<$-2.93	\\
J0324--27	&	114	&	-2.69	&		&		&		&	$<$-1.41	\\
J0329--37	&	48	&	-1.69	&		&		&		&	$<$-3.15	\\
J0336--26	&	50	&	-2.79	&		&		&		&	$<$-2.01	\\
J0339--24	&	60	&	-2.52	&		&		&		&	$<$-2.13	\\
J0341--22	&	77	&	-0.93	&		&		&		&	$<$-3.49	\\
J0351--51	&	53	&	-1.60	&		&		&		&	$<$-3.16	\\
J0356--39	&	56	&	-1.53	&		&		&		&	$<$-3.19	\\
J0405--40	&	52	&	-1.29	&		&		&		&	$<$-3.48	\\
J0413--44	&	64	&	-1.49	&		&		&		&	$<$-3.11	\\
J0427--24	&	63	&	-2.62	&		&		&		&	$<$-1.99	\\
J0427--33	&	72	&	-1.58	&		&		&		&	$<$-2.92	\\
\hline
\end{tabular}
\end{center}
\caption{ROSAT X-ray counterparts to Table~\ref{table:th1} sources along with estimated $L_X/L_{bol}$. For objects without X-ray detection, we adopt the ROSAT All-Sky Survey limit of $2\times10^{-13}$~ergs~cm$^{-2}$~s$^{-1}$ provided in \citet{Schmitt:1995}. Distances listed are the Tuc-Hor kinematic estimates, but note that $L_X/L_{bol}$ is independent of distance. Note that J0242--53E and J0242--53W match the same X-ray source. } \label{xray1}
\end{table}
%\end{longtable}

\clearpage

\begin{table}
\begin{center}
\begin{tabular}{lrc}
\hline
  \multicolumn{1}{c}{Criteria} &
  \multicolumn{1}{c}{Number} & 
  \multicolumn{1}{c}{Percent} \\
\hline
  This work & 20/58 & $34^{+7}_{-6}$ \\
  Rodriguez et al.\ (2011) & 14/54 & $26^{+7}_{_5}$ \\
  \hline
  $<$M4 & 10/17 & $58^{+10}_{-12}$ \\
  $\geq$M4 & 10/41 & $24^{+8}_{-5}$ \\
  \hline
  Within 60pc, $<$M4 & 6/11 & $54^{+13}_{-14}$ \\
  Within 60pc, $\geq$M4 & 8/24 & $33^{+11}_{-8}$ \\
\hline\end{tabular}
\end{center}
\caption{ROSAT X-ray detection rates. Distance are estimated from the convergent point method assuming Tuc-Hor membership. Stars with distances $>$60~pc may not be members of Tuc-Hor (see Section~\ref{banyan}). Binomial errors are quoted for our percentages \citep{Burgasser:2003}. }\label{xray2}
\end{table}

%\clearpage

% Distance comparisons
\begin{table}
\begin{center}
\begin{tabular}{lccc}
\hline
  \multicolumn{1}{c}{Name} &
  \multicolumn{1}{c}{Parallax} &
  \multicolumn{1}{c}{Conv.\ Point} &
  \multicolumn{1}{c}{BANYAN} \\
   \multicolumn{1}{c}{} &
  \multicolumn{1}{c}{D (pc)} &
  \multicolumn{1}{c}{D (pc)} &
  \multicolumn{1}{c}{D (pc)} \\
\hline
  HD 105 & 40.0 & 39.5 & 39.0\\
  HD 987 & 44.0 & 47.9 & 47.5\\
  HD 1466 & 41.0 & 44.0 & 43.5\\
  HD 2884 & 43.0 & 45.5 & 44.5\\
  HD 3003 & 47.0 & 46.4 & 45.5\\
  HD 3221 & 46.0 & 46.4 & 45.0\\
  HD 8558 & 49.0 & 46.9 & 45.0\\
  CC Phe & 37.0 & 41.3 & 40.5\\
  DK Cet & 42.0 & 41.7 & 40.5\\
  HD 13183 & 50.0 & 51.2 & 47.5\\
  HD 13246 & 45.0 & 46.6 & 44.5\\
  phi  Eri & 47.0 & 47.9 & 45.0\\
  epsilon  Hyi & 47.0 & 49.3 & 47.5\\
  HD 22705 & 42.0 & 45.6 & 44.0\\
  HD 29615 & 55.0 & 59.3 & 53.5\\
  HD 30051 & 58.0 & 68.7 & 59.0\\
  HD 32195 & 60.0 & 65.1 & 59.5\\
  alpha  Pav & 56.0 & 56.4 & 57.0\\
  HD 202917 & 46.0 & 48.9 & 50.0\\
  HD 207575 & 45.0 & 48.2 & 48.5\\
  HD 207964 & 47.0 & 48.4 & 48.5\\
  DS Tuc & 46.0 & 40.3 & 40.5\\
\hline\end{tabular}
\caption{Comparison of distances for known Tuc-Hor members with parallaxes \citep{Torres:2008} against our convergent point kinematic distance and those distances predicted by BANYAN \citep{Malo:2012}. 
} \label{banyan2}
\end{center}
\end{table}

%\clearpage

% UVW information
\begin{table}
\begin{center}
\begin{tabular}{lcrrrr}
\hline
  \multicolumn{1}{c}{ID} &
  \multicolumn{1}{c}{D (pc)} & 
  \multicolumn{1}{c}{RV (km/s)} & 
  \multicolumn{1}{c}{U (km/s)} & 
  \multicolumn{1}{c}{V (km/s)} & 
  \multicolumn{1}{c}{W (km/s)} \\
\hline
J0202--31	&	60$^a$	& $	16.7	\pm	1.5	$ & $-14.2\pm2.4$	 & $-24.2\pm4.2$	 & $-9.3\pm2.0$ \\
J0220--58	&	48	& $	7.4	\pm	1.5	$ & $-12.2\pm2.7$ & 	$-20.2\pm3.3$	 & $2.9\pm2.2$\\
J0259--42	&	100$^a$	& $	15.3	\pm	1.5	$ & $-10.7\pm2.7$	 & $-21.8\pm3.5$	 & $-3.7\pm2.5$ \\
J0315--53	&	49	& $	9.4	\pm	1.5	$ & $-8.1\pm2.4$ & $-19.3\pm3.3$ & 	$3.1\pm2.7$ \\
J0413--44	&	56	& $	2.3	\pm	6.6	$ & $-5.9\pm2.0$ & 	$-11.0\pm4.7$	 & $8.6\pm5.3$ \\
\hline
J0236--52	&	42	& $	16	\pm	1	$ & $-12.9\pm2.8$	 & $-21.6\pm2.9$	 & $-5.7\pm1.9$ \\
		&	25$^b$	& 	& $-7.7\pm1.9$	 & $-16.3\pm2.0$	 & $-8.9\pm1.4$ \\
\hline\end{tabular}
\end{center}
\caption{UVW velocities for stars with measured radial velocities (see Appendix~\ref{notesys} for more details). Distances have uncertainties of 20\%. The average UVW for Tuc-Hor members is $-9.9\pm1.5$, $-20.9\pm0.8$, $-1.4\pm0.9$~km/s \citep{Torres:2008}.
\newline $^a$: Kinematic distance assuming membership in the Columba Association. The average UVW velocity for Columba members is $-13.2\pm1.3$, $-21.8\pm0.8$, $-5.9\pm1.2$~km/s \citep{Torres:2008}.
\newline $^b$: Isochrone distance from \citet{Torres:2000} and \citet{ZS04}. } \label{tab:uvw}
\end{table}

\clearpage

% Summary table
\begin{table}
\begin{center}
\begin{tabular}{cccccrrrr}

\hline
WISE 		 & Group 	& Ks 	& R--Ks 	& Sp.  	& \multicolumn{2}{c}{Conv. Point} & \multicolumn{2}{c}{BANYAN} \\
Designation &  		& (mag)	& (mag) 	& Type 	& Prob.\	& Dist.\	& Prob.\	& Dist.\ \\
\hline

\hline
J012758.87-603224.5	&	TH	&	10.2	&	4.4	&	M4.0	&	96	&	49	&	98	&	47	\\
J015057.01-584403.4	&	TH	&	8.6	&	4.0	&	M2.9	&	95	&	48	&	97	&	46	\\
J015325.09-683322.8	&	TH	&	10.2	&	4.9	&	M5.1	&	79	&	45	&	99	&	44	\\
J020020.08-661402.0	&	TH	&	9.9	&	4.7	&	M4.3	&	79	&	52	&	98	&	50	\\
J020701.85-440638.3	&	TH	&	8.4	&	3.4	&	M3.4	&	99	&	46	&	98	&	44	\\
J021258.28-585118.3	&	TH	&	8.4	&	3.6	&	M2.1	&	95	&	50	&	96	&	47	\\
J021533.37-562717.6	&	TH	&	11.0	&	5.3	&	M5.4	&	94	&	50	&	95	&	47	\\
J022051.50-582341.3	&	TH	&	8.8	&	4.2	&	M3.0	&	78	&	45	&	96	&	44	\\
J022244.32-602247.7	&	TH	&	8.1	&	4.5	&	M4.0	&	58	&	32	&	65	&	33	\\
J022424.69-703321.2	&	TH	&	9.5	&	4.7	&	M4.3	&	98	&	47	&	100	&	46	\\
J023219.44-574611.9	&	TH	&	10.2	&	5.4	&	M4.4	&	66	&	51	&	95	&	48	\\
J024127.29-304915.1	&	TH	&	10.3	&	4.6	&	M4.7	&	70	&	45	&	59	&	43	\\
J024202.14-535914.7	&	TH	&	10.0	&	4.7	&	M4.6	&	55	&	44	&	92	&	42	\\
J024204.15-535900.0	&	TH	&	9.3	&	4.7	&	M4.7	&	88	&	44	&	96	&	43	\\
J024746.49-580427.4	&	TH	&	8.4	&	3.9	&	M3.0	&	92	&	45	&	95	&	44	\\
J025022.35-654555.2	&	TH	&	9.4	&	3.8	&	M3.7	&	85	&	56	&	97	&	52	\\
J025059.67-340905.3	&	TH	&	9.6	&	4.2	&	M4.5	&	70	&	49	&	58	&	46	\\
J025433.25-510831.4	&	TH	&	7.8	&	3.3	&	M1.5	&	100	&	46	&	97	&	44	\\
J025531.87-570252.3	&	TH	&	10.2	&	4.8	&	M4.9	&	93	&	47	&	95	&	45	\\
J030505.65-531718.4	&	TH	&	10.3	&	5.2	&	M5.4	&	77	&	45	&	93	&	44	\\
J031049.48-361647.3	&	TH	&	9.8	&	4.8	&	M4.3	&	49	&	45	&	96	&	43	\\
J031145.52-471950.2	&	TH	&	9.6	&	4.0	&	M4.3	&	52	&	47	&	92	&	44	\\
J031523.72-534253.9	&	TH	&	10.4	&	5.4	&	M5.2	&	64	&	51	&	87	&	47	\\
J032440.63-390422.8	&	TH	&	9.0	&	4.1	&	M4.2	&	89	&	47	&	94	&	44	\\
J032916.57-370250.2	&	TH	&	9.8	&	5.0	&	M4.3	&	52	&	48	&	95	&	45	\\
J035122.95-515458.1	&	TH	&	9.8	&	4.3	&	M4.2	&	98	&	53	&	93	&	49	\\
J035616.31-391521.8	&	TH	&	9.6	&	5.0	&	M5.0	&	98	&	56	&	65	&	50	\\
J040539.68-401410.5	&	TH	&	9.0	&	4.4	&	M4.2	&	87	&	52	&	69	&	48	\\
J041336.14-441332.4	&	TH	&	9.9	&	4.1	&	M3.9	&	97	&	64	&	81	&	54	\\
\hline
J014246.89-512646.9	&	Col	&	10.1	&	3.7	&	M6.5	&	89	&	72	&	73	&	66	\\
J023359.89-181152.5	&	Col	&	9.2	&	3.8	&	M3.7	&	85	&	85	&	97	&	77	\\
J025901.49-423220.4	&	Col	&	11.4	&	4.4	&	M4.2	&	91	&	106	&	69	&	92	\\
J030509.79-372505.8	&	Col	&	8.7	&	3.8	&	M1.9	&	65	&	81	&	95	&	73	\\
J030839.55-384436.3	&	Col	&	10.4	&	4.8	&	M4.2	&	100	&	60	&	72	&	56	\\
J032144.76-330949.5	&	Col	&	10.4	&	3.9	&	M5.8	&	52	&	96	&	78	&	83	\\
J032443.06-273323.1	&	Col	&	11.7	&	5.0	&	M5.5	&	81	&	112	&	64	&	95	\\
J033901.64-243406.1	&	Col	&	10.0	&	4.8	&	M5.9	&	56	&	58	&	71	&	53	\\
J040711.50-291834.3	&	Col	&	8.2	&	3.2	&	M0.0	&	78	&	81	&	93	&	72	\\
J042726.28-245527.4	&	Col	&	10.8	&	4.2	&	M4.5	&	90	&	55	&	77	&	51	\\
J042745.66-332742.6	&	Col	&	10.4	&	5.1	&	M4.8	&	86	&	63	&	88	&	57	\\
J043138.61-304250.9	&	Col	&	9.3	&	3.5	&	M3.2	&	77	&	90	&	87	&	77	\\
\hline
J020306.68-554542.1	&	ABD	&	10.4	&	4.5	&	M4.5	&	100	&	73	&	69	&	69	\\
J021053.50-460351.4$^a$&	ABD	&	10.3	&	3.8	&	M4.2&	\nodata & \nodata & \nodata & \nodata \\
\hline
\end{tabular}
\end{center}
\caption{Final table of candidates for Table~\ref{table:th1} sources after eliminating likely field contaminants. 
Candidates are listed by moving group (TH=Tuc-Hor, Col=Columba, and ABD=AB~Dor) and sorted by RA. Distances and membership probabilities are listed for the particular group. Spectral types are estimated from the TiO5 index or J--W2 color. 
R magnitudes come from NOMAD which in turn are drawn from USNO-B1 and UCAC2; an error of 0.3 magnitudes is assumed.
In Appendix~\ref{notesys}, we discuss several of these candidate young systems in more detail, highlighting those that remain unconfirmed or doubtful members of the listed groups.
Further work is needed to fully confirm these as members of these groups.
\\Notes: $^a$ J0210--46 is a low-mass companion to AB Dor member CD-46~644; see Appendix~\ref{notesys}.
} \label{tab:final}
%\end{longtable}
\end{table}

\clearpage

% J-W2 figure

% PM figure2

% PM figure

\clearpage

% Sp Type figure (relationship)

% WISE ISO figure

% Sp Type figure (distribution)

% THPM figure

% TH CMD figure

% Wifes spectra figure
\begin{figure}[htb]
\begin{center}
\includegraphics[width=13cm,angle=0]{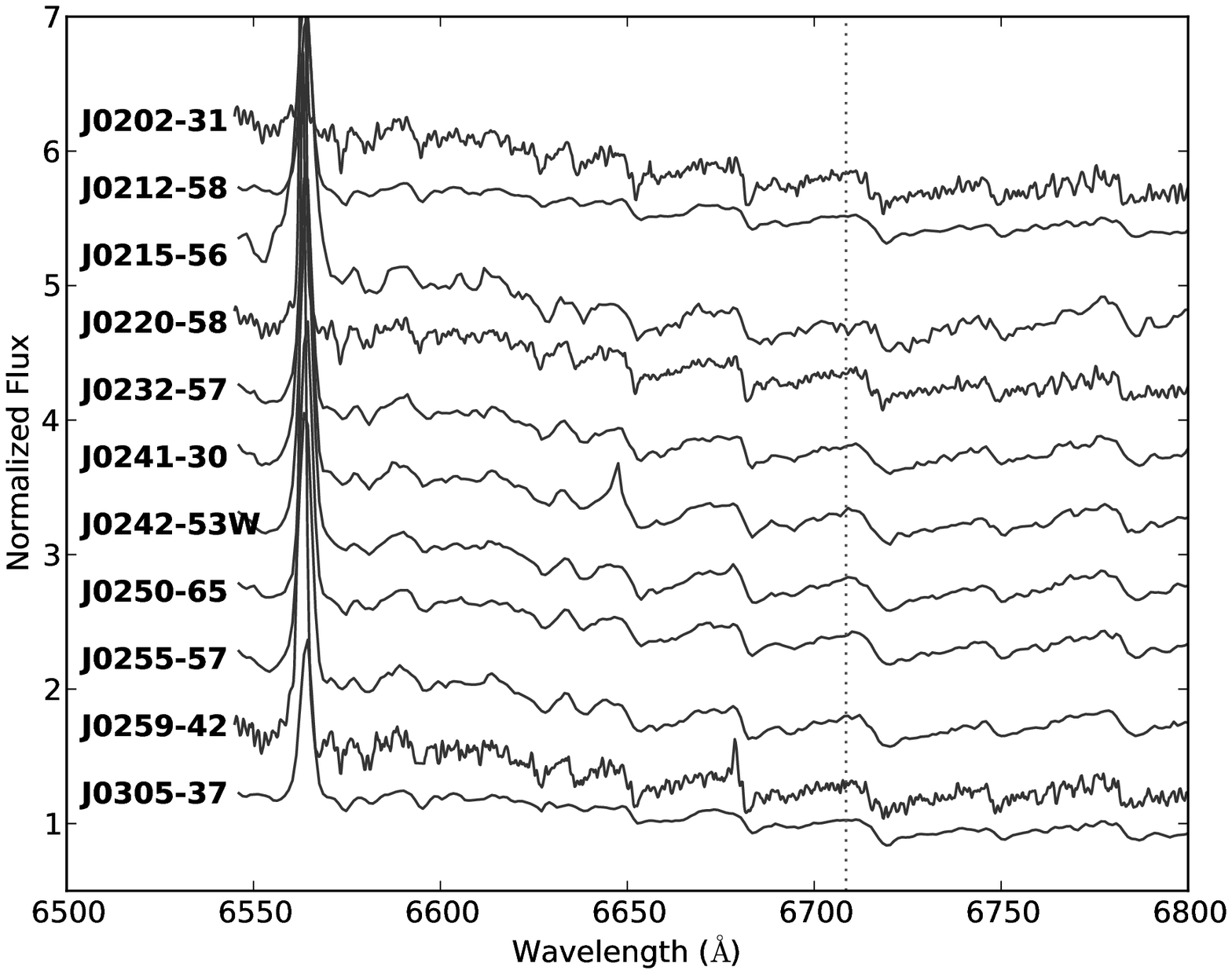}
\includegraphics[width=13cm,angle=0]{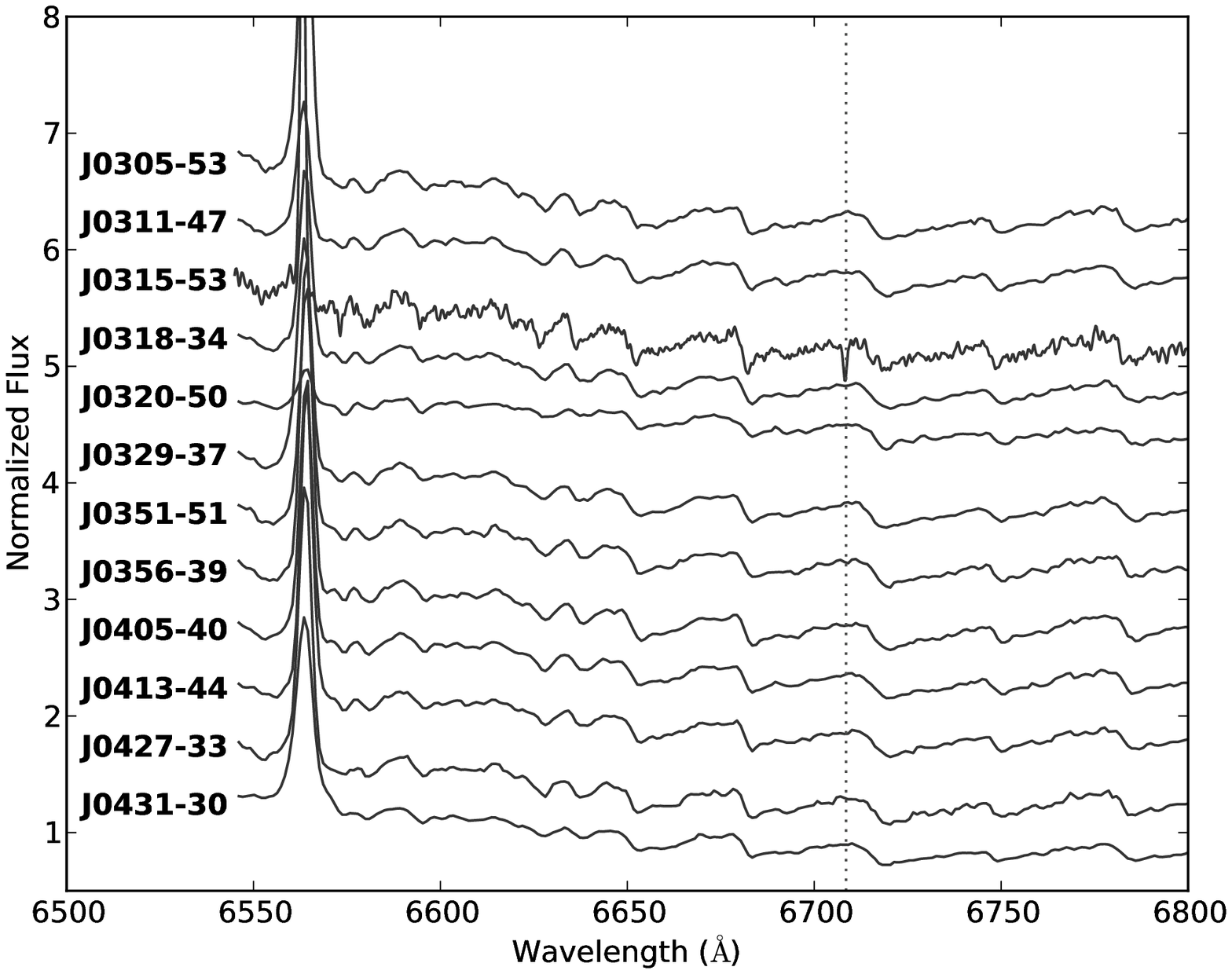}
\end{center}
\caption{WiFeS spectra for our candidate objects. 
Spectra for J0202--31, J0220--58, J0259--42, and J0315--53 have R$\sim$7000, all others have R$\sim$3000. The vertical line indicates the location of 
the Li~6708\AA\, absorption feature; only J0315--53 shows strong Li absorption. 
The feature near 6650\AA\, for J0241--30 and 6680\AA\, for J0259--42 is due to uncorrected cosmic ray hits.
J0202--31 is a binary system.}
\label{wifes}
\end{figure}

%TiO5 vs J-W2 spectral types figure
\begin{figure}[htb]
\begin{center}
\includegraphics[width=14cm,angle=0]{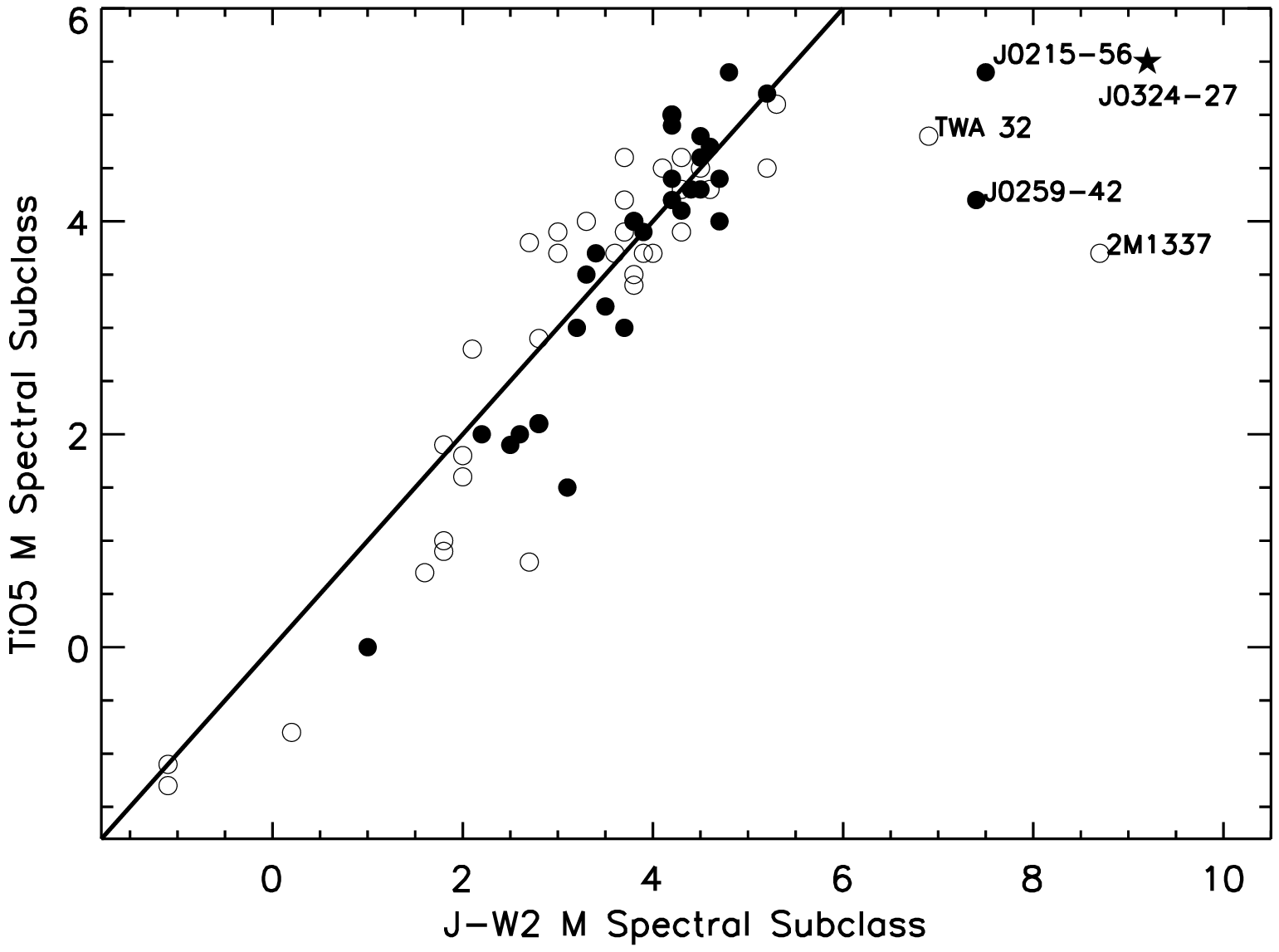}
\end{center}
\caption{Comparison of M-dwarf spectral subtypes determined from J--W2 color (see Section~\ref{spectypes}) 
and the TiO5 index from optical spectra \citep{Reid:1995}. Filled circles are from Table~\ref{table:th1}. 
Open circles are candidate young stars published by \citet{Rodriguez:2011}. 
J0324--27 is labeled with a star symbol to indicate that the spectral type is estimated from the near-IR spectrum (Section~\ref{thspex1}).
J0259--42 and J0324--27 displays clear signs of excess IR emission, while J0215--56 is more ambiguous (see Figure~\ref{fig:sed}). 
The IR excesses around TWA~32 and 2M1337 have been previously noted \citep{Schneider:2012a,Schneider:2012b}.
}
\label{sptype}
\end{figure}

% IRTF spectra figure
\begin{figure}[htb]
\begin{center}
\includegraphics[width=14cm,angle=0]{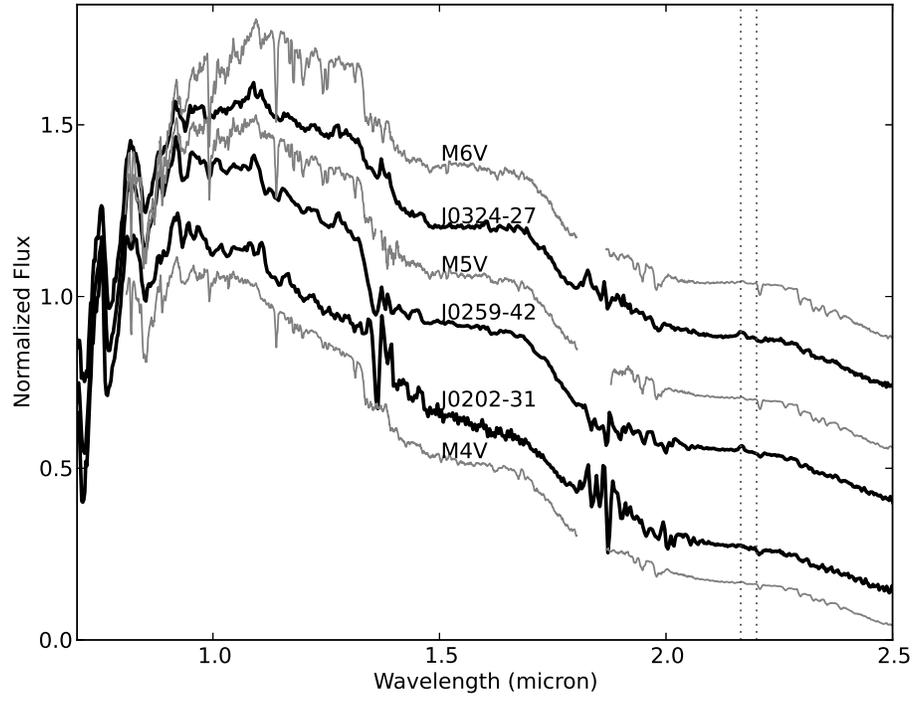}
\end{center}
\caption{SpeX R$\sim$100 spectra for three of our Tuc-Hor candidates. 
Thin, grey spectra denote M6V, M5V, and M4V comparison field dwarfs from the IRTF SpeX Library.
The vertical dotted lines indicate the location of the Br-$\gamma$ and Na~I features. J0259--42 and J0324--27 both show Br-$\gamma$ emission and relatively weak Na~I absorption. J0202--31 is a binary system.
}
\label{th_irtf}
\end{figure}

% Halpha figure
\begin{figure}[htb]
\begin{center}
\includegraphics[width=14cm,angle=0]{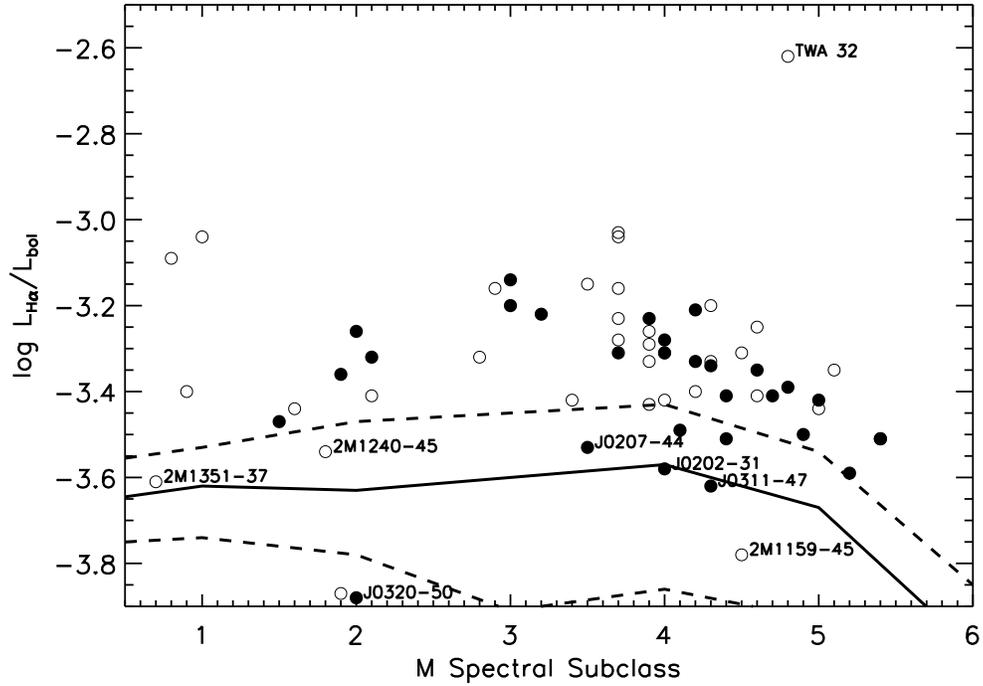}
\end{center}
\caption{$L_{H\alpha} / L_{bol}$ for the Table~\ref{table:th1} Tuc-Hor candidates for which we have measured H$\alpha$ equivalent widths. The solid line indicates the average $L_{H\alpha} / L_{bol}$ for the field population studied in \citet{West:2004} and its uncertainty (dashed lines) which includes the 1$\sigma$ distribution of $L_{H\alpha} / L_{bol}$ (see Figure 5 in \citealt{West:2004}). 
Open circles denote $\sim$10--20~Myr-old TWA and Scorpius-Centaurus candidates from \citet{Rodriguez:2011} analyzed in the same way. Some outliers in the figure are labeled. 
}
\label{halpha}
\end{figure}

% Na 8200 spectrum figure
\begin{figure}[htb]
\begin{center}
\includegraphics[width=14cm,angle=0]{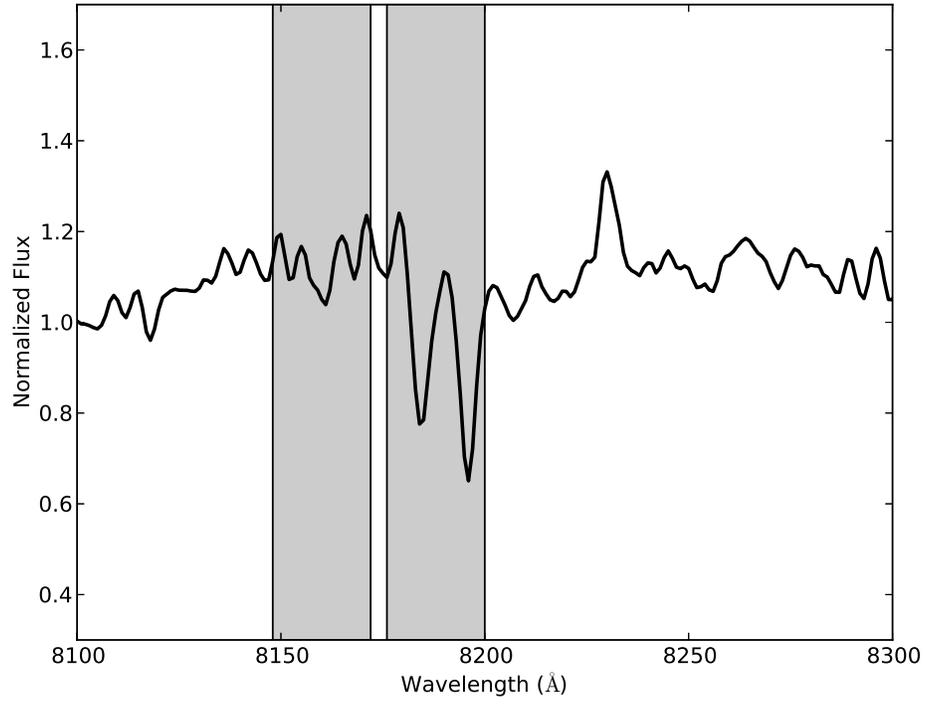}
\end{center}
\caption{Spectrum for J0259--42, showing the region around the sodium doublet at 8183 and 8195\AA. Highlighted in grey are the two regions used to compute the Na~I index (see Section~\ref{sodium} and \citealt{Lawson:2009}).
}
\label{naspec}
\end{figure}

% Na Index figure
\begin{figure}[htb]
\begin{center}
\includegraphics[width=14cm,angle=0]{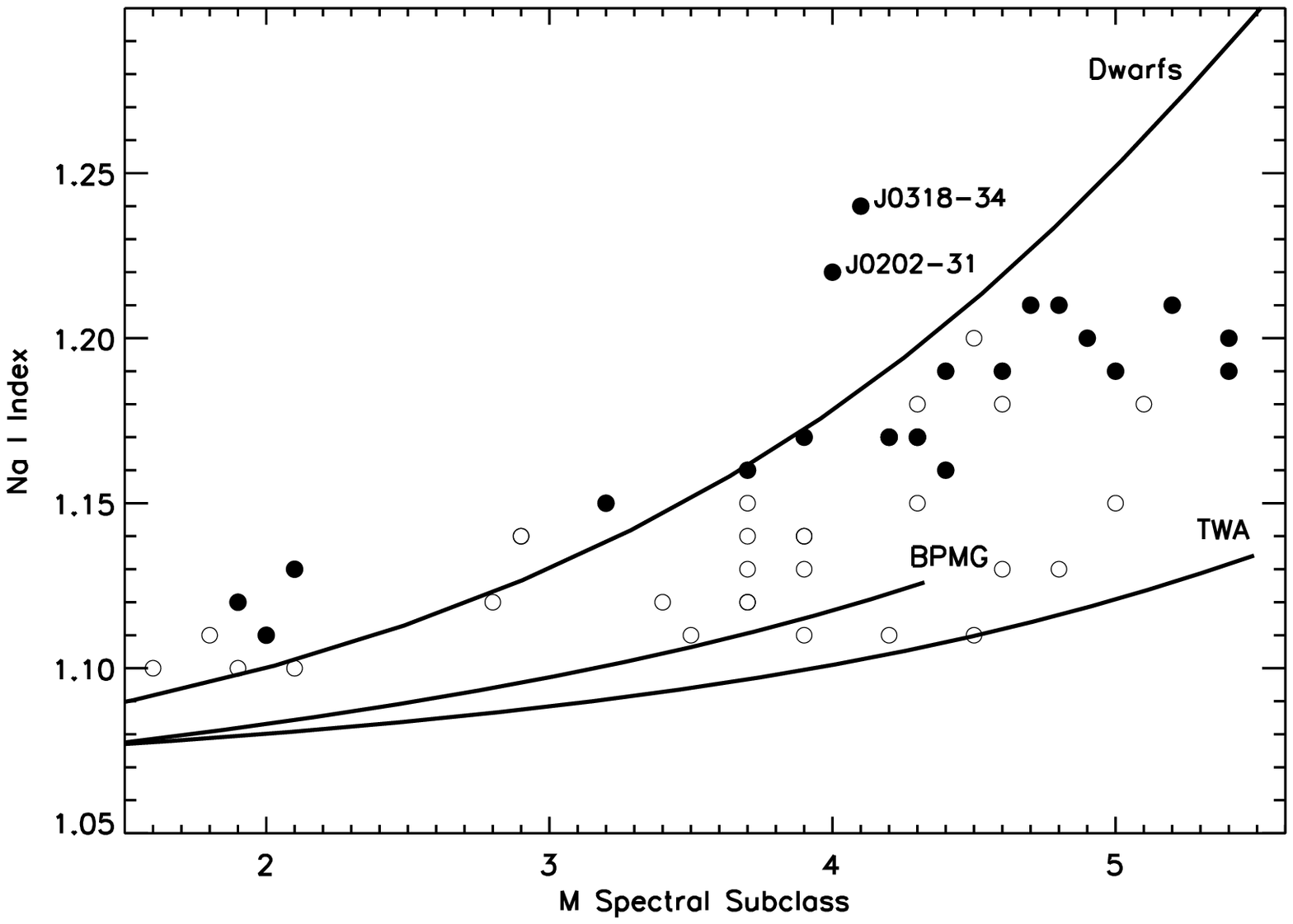}
\end{center}
\caption{Na~I index values (see \citealt{Lawson:2009} and Section~\ref{sodium}) for the Table~\ref{table:th1} Tuc-Hor candidates for which we have measured Na~I (filled circles). 
The Na~I index curve for TWA beyond M4.5 is not well constrained as only a single TWA member is used there (see Figure~1 in \citealt{Lawson:2009}).
Open circles denote $\sim$10--20~Myr-old TWA and Scorpius-Centaurus candidates from \citet{Rodriguez:2011} analyzed in the same way.
For spectral types earlier than M4, the Na~I index is not a reliable means to distinguish between older field dwarfs and young ($>$10~Myr) moving group members.
The Na~I indices of most of the Tuc-Hor candidates later than M4 appear consistent with ages intermediate between that of TWA or the $\beta$~Pic moving group (which have ages $\sim$10~Myr) and the field dwarf population ($\gtrsim$1~Gyr).
}
\label{naindex}
\end{figure}

% Xray figure
\begin{figure}[htb]
\begin{center}
\includegraphics[width=14cm,angle=0]{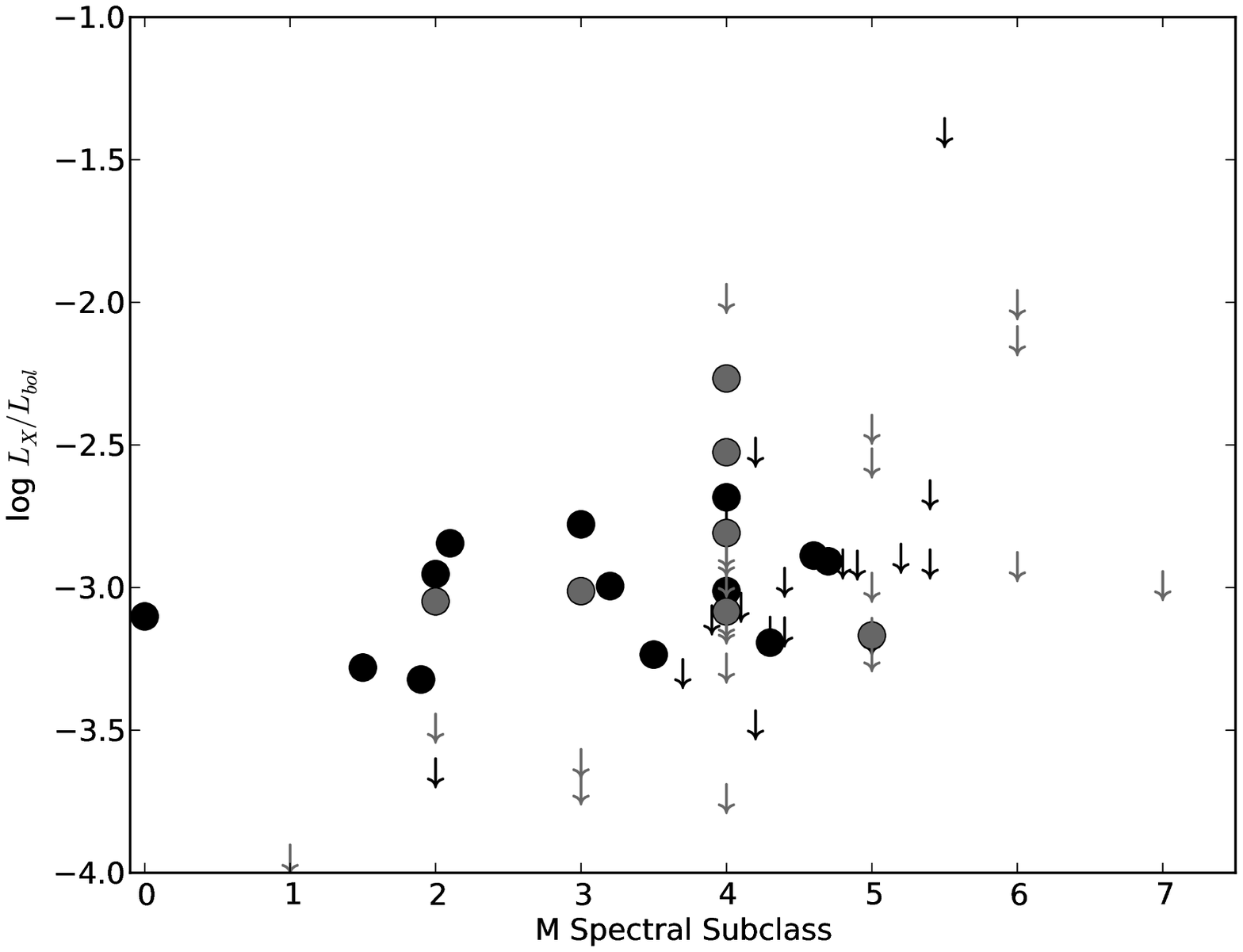}
\end{center}
\caption{X-ray detections and limits for the Table~\ref{table:th1} Tuc-Hor candidates. Dark grey symbols denote objects with spectral types estimated by their J--W2 colors. Downward arrows denote upper limits as described in Section~\ref{xraysec}.
}
\label{fig:xray1}
\end{figure}

% XYZ figure
\begin{figure}[htb]
\begin{center}
\includegraphics[width=14cm,angle=0]{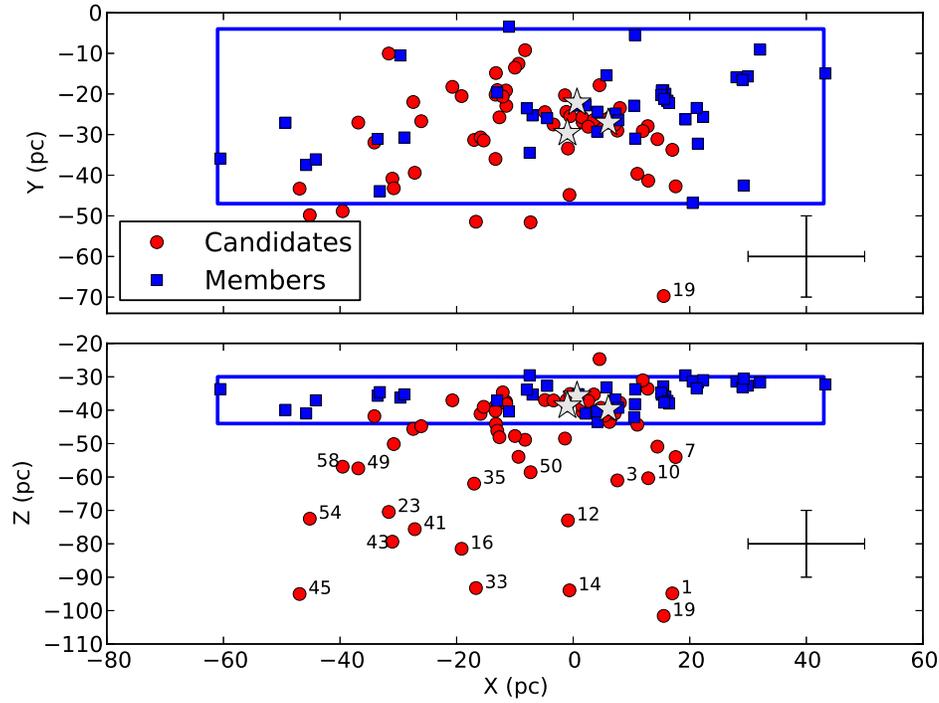}
\end{center}
\caption{XYZ positions for candidates, adopting kinematic distances (see Section~\ref{cpanalysis}), compared to known members as drawn from \citet{Torres:2008}. XYZ is defined in the same fashion as UVW, with X positive towards the Galactic center, Y in the direction of Galactic rotation, and Z positive towards the North Galactic Pole. 
The box outlines the spread of Tuc-Hor members \citep{Torres:2008}. A 10-pc error bar is also shown, which corresponds to a 20\% distance uncertainty at the typical distance of 50~pc. 
Prominent outliers are labeled with their Table~\ref{table:th1} index and are discussed in Appendix~\ref{notesys}.
The three star symbols represent the three candidate Tuc-Hor members listed in Table~\ref{tab:uvw} and shown in Figure~\ref{fig:uvw}. 
}
\label{fig:xyz}
\end{figure}

% UVW figure
\begin{figure}[htb]
\begin{center}
\includegraphics[width=14cm,angle=0]{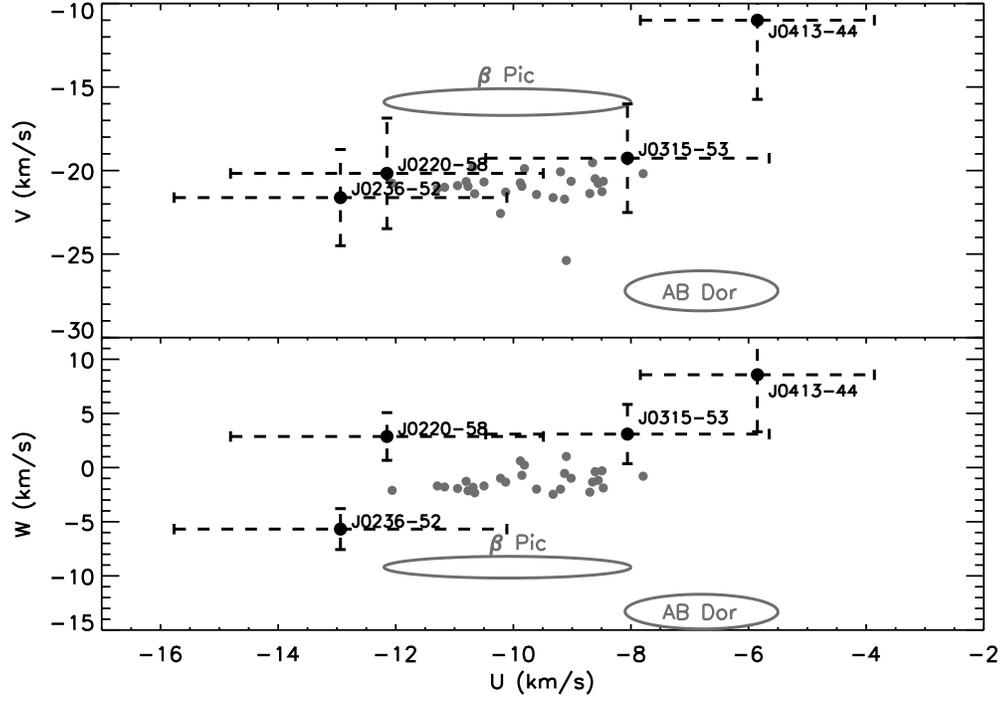}
\end{center}
\caption{UVW velocities for candidates (see Table~\ref{tab:uvw} and Appendix~\ref{notesys}) with measured radial velocities compared to several young moving groups \citep{Torres:2008}. 
A distance of 42~pc is used for J0236--52. 
J0220--58, J0236--52, and J0315--53 have XYZ consistent with Tuc-Hor membership as shown in Figure~\ref{fig:xyz}.
The small grey circles correspond to members of Tuc-Hor \citep{Torres:2006,Torres:2008}.
}
\label{fig:uvw}
\end{figure}

%SED figure
\begin{figure}[htb]
\begin{center}
\includegraphics[width=14cm,angle=0]{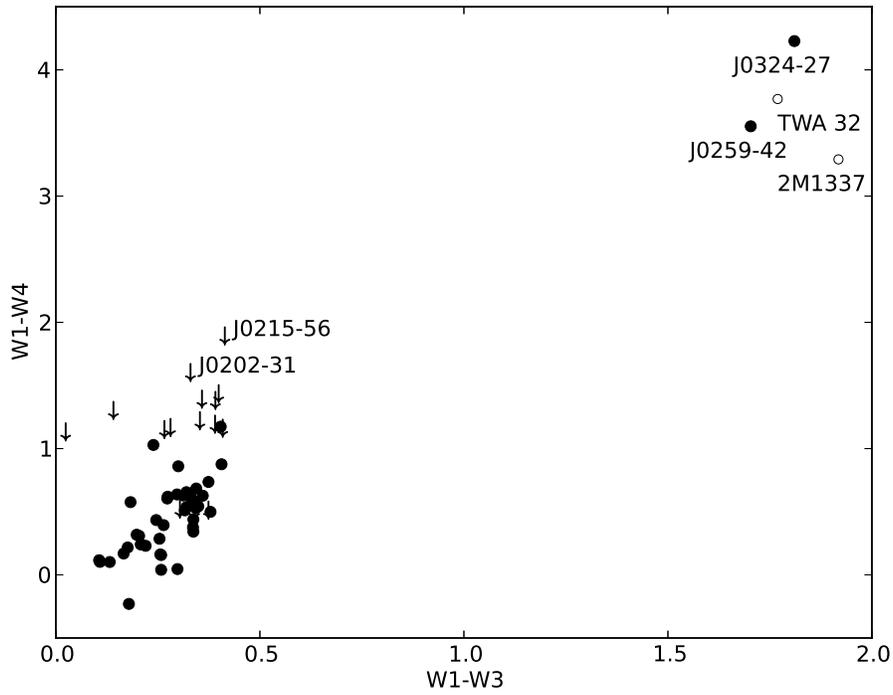}
\end{center}
\caption{WISE $W1-W3$ and $W1-W4$ colors for the Table~\ref{table:th1} Tuc-Hor candidates. Downward arrows denote systems for which upper limits at $W4$ are available. J0259--42 and J0324--27 have clear infrared excesses, suggesting the possibility that warm (T$\sim$300~K) dust orbits in a disk within these systems (see Appendix~\ref{notesys}). 
For comparison, the dusty systems in the sample of \citet{Rodriguez:2011} are also shown and labeled \citep[see also][]{Schneider:2012a,Schneider:2012b}.
}
\label{fig:sed}
\end{figure}


\begin{thebibliography}

\bibitem[Allers et al.(2007)]{Allers:2007} Allers, K.~N., Jaffe, D.~T., Luhman, K.~L., et al.\ 2007, \apj, 657, 511
\bibitem[Barrado y Navascu{\'e}s et al.(2001)]{Barrado:2001} Barrado y Navascu{\'e}s, D., Stauffer, J.~R., Brice{\~n}o, C., et al.\ 2001, \apjs, 134, 103 
\bibitem[Baraffe et al.(1998)]{Baraffe:1998} Baraffe, I., Chabrier, G., Allard, F., \& Hauschildt, P.~H.\ 1998, \aap, 337, 403
\bibitem[Bochanski et al.(2010)]{Bochanski:2010} Bochanski, J.~J., Hawley, S.~L., Covey, K.~R., et al.\ 2010, \aj, 139, 2679 
\bibitem[Burgasser et al.(2003)]{Burgasser:2003} Burgasser, A.~J., Kirkpatrick, J.~D., Reid, I.~N., et al.\ 2003, \apj, 586, 512
\bibitem[Burrows et al.(1997)]{Burrows:1997} Burrows, A., Marley, M., Hubbard, W.~B., et al.\ 1997, \apj, 491, 856
\bibitem[Cardelli et al.(1989)]{Cardelli:1989} Cardelli, J.~A., Clayton, G.~C., \& Mathis, J.~S.\ 1989, \apj, 345, 245 
\bibitem[Chauvin et al.(2004)]{Chauvin:2004} Chauvin, G., Lagrange, A.-M., Dumas, C. et al 2004, A\&A 425, L29
\bibitem[Chauvin et al.(2005)]{Chauvin:2005} Chauvin, G., Lagrange, A.-M., Dumas, C. et al 2005, A\&A 438, L25
\bibitem[Chauvin et al.(2010)]{Chauvin:2010} Chauvin, G., Lagrange, A.-M., Bonavita, M., et al.\ 2010, \aap, 509, A52 
\bibitem[Close et al.(2003)]{Close:2003} Close, L.~M., Siegler, N., Freed, M., \& Biller, B.\ 2003, \apj, 587, 407
\bibitem[Close et al.(2007)]{Close:2007} Close, L.~M., Zuckerman, B., Song, I., et al.\ 2007, \apj, 660, 1492 
\bibitem[Cutri et al.(2003)]{Cutri:2003} Cutri, R.~M., et al.\ 2003, The IRSA 2MASS All-Sky Point Source Catalog, NASA/IPAC Infrared Science Archive.~http://irsa.ipac.caltech.edu/applications/Gator/
\bibitem[Cushing et al.(2004)]{Cushing:2004} Cushing, M.~C., Vacca, W.~D., \& Rayner, J.~T.\ 2004, \pasp, 116, 362 %SpeXtool
\bibitem[de Bruijne(1999)]{deBruijne:1999} de Bruijne, J.~H.~J.\ 1999, \mnras, 306, 381 
\bibitem[DENIS Consortium(2005)]{DENIS} The DENIS Consortium.\ 2005, VizieR Online Data Catalog, 2263, 0 
\bibitem[Dopita et al.(2007)]{Dopita:2007} Dopita, M., Hart, J., McGregor, P., Oates, P., Bloxham, G., \& Jones, D.\ 2007, \apss, 310, 255 %wifes
\bibitem[Dupuy \& Liu(2012)]{Dupuy:2012} Dupuy, T.~J., \& Liu, M.~C.\ 2012, \apjs, 201, 19
\bibitem[Faherty et al.(2011)]{Faherty:2011} Faherty, J.~K., Burgasser, A.~J., Bochanski, J.~J., et al.\ 2011, \aj, 141, 71
\bibitem[Faherty et al.(2012)]{Faherty:2012} Faherty, J.~K., Burgasser, A.~J., Walter, F.~M., et al.\ 2012, \apj, 752, 56 
\bibitem[Faherty et al.(2013)]{Faherty:2013} Faherty, J.~K., Rice, E.~L., Cruz, K.~L., Mamajek, E.~E., \& N{\'u}{\~n}ez, A.\ 2013, \aj, 145, 2
\bibitem[Farihi et al.(2005)]{Farihi:2005} Farihi, J., Becklin, E.~E., \& Zuckerman, B.\ 2005, \apjs, 161, 394 
\bibitem[Findeisen \& Hillenbrand(2010)]{Findeisen:2010} Findeisen, K., \& Hillenbrand, L.\ 2010, \aj, 139, 1338
\bibitem[Findeisen et al.(2011)]{Findeisen:2011} Findeisen, K., Hillenbrand, L., \& Soderblom, D.\ 2011, \aj, 142, 23
\bibitem[Hambly et al.(2001)]{Hambly:2001} Hambly, N.~C., MacGillivray, H.~T., Read, M.~A., et al.\ 2001, \mnras, 326, 1279 % SuperCosmos
\bibitem[Hauschildt et al.(1999)]{Hauschildt:1999} Hauschildt, P.~H., Allard, F., \& Baron, E.\ 1999, \apj, 512, 377 % NextGen
\bibitem[Henry et al.(2006)]{Henry:2006} Henry, T.~J., Jao, W.-C., Subasavage, J.~P., et al.\ 2006, \aj, 132, 2360 %RECONS paper 17
\bibitem[Hughes et al.(2008)]{Hughes:2008} Hughes, A.M., Wilner, D.J., Kamp, I., \& Hogerheijde, M.R.\ 2008, \apj, 681, 626
\bibitem[Jones(1971)]{Jones:1971} Jones, D.~H.~P.\ 1971, \mnras, 152, 231 
\bibitem[Kastner et al.(1997)]{Kastner:1997} Kastner, J.~H., Zuckerman, B., Weintraub, D.~A., \& Forveille, T.\ 1997, Science, 277, 67
\bibitem[Kastner et al.(2002)]{Kastner:2002} Kastner, J.~H., Huenemoerder, D.~P., Schulz, N.~S., Canizares, C.~R., \& Weintraub, D.~A.\ 2002, \apj, 567, 434 %TW Hya accretion, xrays
\bibitem[Kastner et al.(2008)]{Kastner:2008} Kastner, J.H., Zuckerman, B., Hily-Blant, P., \& Forveille, T.\ 2008, \aap, 492, 469 % V4046 Sgr
\bibitem[Kenyon \& Hartmann(1995)]{KH:1995} Kenyon, S.~J., \& Hartmann, L.\ 1995, \apjs, 101, 117 
\bibitem[Kirkpatrick et al.(2008)]{Kirkpatrick:2008} Kirkpatrick, J.~D., Cruz, K.~L., Barman, T.~S., et al.\ 2008, \apj, 689, 1295
\bibitem[Kirkpatrick et al.(2011)]{Kirkpatrick:2011} Kirkpatrick, J.~D., Cushing, M.~C., Gelino, C.~R., et al.\ 2011, \apjs, 197, 19 
\bibitem[Kraus \& Hillenbrand(2007)]{Kraus:2007} Kraus, A.~L., \& Hillenbrand, L.~A.\ 2007, \aj, 134, 2340 
\bibitem[Lagrange et al.(2010)]{Lagrange:2010} Lagrange, A.-M., Bonnefoy, M., Chauvin, G., et al.\ 2010, Science, 329, 57
\bibitem[Lawson et al.(2009)]{Lawson:2009} Lawson, W.~A., Lyo, A.-R., \& Bessell, M.~S.\ 2009, \mnras, 400, L29
\bibitem[Linsky et al.(2001)]{Linsky:2001} Linsky, J., Redfield, S., Ayres, T., Brown, A., \& Harper, G.\ 
2001, "Eta Carinae and Other Mysterious Stars: The Hidden Opportunities of Emission Spectroscopy," ASP 
Conf. Procedings 242, 247
\bibitem[Looper et al.(2010)]{Looper:2010b} Looper, D.~L., Bochanski, J.~J., Burgasser, A.~J., et al.\ 2010, \aj, 140, 1486 
\bibitem[Lucas et al.(2001)]{Lucas:2001} Lucas, P.~W., Roche, P.~F., Allard, F., \& Hauschildt, P.~H.\ 2001, \mnras, 326, 695
\bibitem[Lyo et al.(2004)]{Lyo:2004} Lyo, A.-R., Lawson, W.~A., \& Bessell, M.~S.\ 2004, \mnras, 355, 363
\bibitem[Malo et al.(2013)]{Malo:2012} Malo, L., Doyon, R., Lafreni{\`e}re, D., et al.\ 2013, \apj, 762, 88 
\bibitem[Mamajek(2005)]{Mamajek:2005} Mamajek, E.~E.\ 2005, \apj, 634, 1385  
\bibitem[Marois et al.(2008)]{Marois:2008} Marois, C., Macintosh, B., Barman, T., et al.\ 2008, Science, 322, 1348 
\bibitem[Marois et al.(2010)]{Marois:2010} Marois, C., Zuckerman, B., Konopacky, Q.~M., Macintosh, B., \& Barman, T.\ 2010, \nat, 468, 1080 
\bibitem[Martin et al.(2005)]{Martin:2005} Martin, D.~C., et al.\ 2005, \apjl, 619, L1 %GALEX
\bibitem[Mohanty et al.(2004a)]{Mohanty:2004a} Mohanty, S., Basri, G., Jayawardhana, R. et al.\ 2004a, \apj, 609, 854
\bibitem[Mohanty et al.(2004b)]{Mohanty:2004b} Mohanty, S., Jayawardhana, R., \& Basri, G.\ 2004b, \apj, 609, 885
\bibitem[Monet et al.(2003)]{Monet:2003} Monet, D.~G., Levine, S.~E., Canzian, B., et al.\ 2003, \aj, 125, 984 % USNO-B1
\bibitem[Neuhaeuser et al.(1995)]{Neuhaeuser:1995} Neuhaeuser, R., Sterzik, M.~F., Schmitt, J.~H.~M.~M., Wichmann, R., \& Krautter, J.\ 1995, \aap, 297, 391 
\bibitem[Preibisch \& Feigelson(2005)]{Preibisch:2005} Preibisch, T., \& Feigelson, E.~D.\ 2005, \apjs, 160, 390 % evolution of x-ray emission in young stars
\bibitem[Qi et al.(2004)]{Qi:2004} Qi, C., Ho, P.T.P., Wilner, D.J., et al.\ 2004, \apj, 616, L11
\bibitem[Qi et al.(2006)]{Qi:2006} Qi, C., Wilner, D.J., Calvet, N., et al.\ 2006, \apj, 636, L157
\bibitem[Qi et al.(2008)]{Qi:2008} Qi, C., Wilner, D.J., Aikawa, Y., et al.\ 2008, \apj, 681, 1396
\bibitem[Rayner et al.(2003)]{Rayner:2003} Rayner, J.~T., Toomey, D.~W., Onaka, P.~M., et al.\ 2003, \pasp, 115, 362 % SpeX
\bibitem[Reid et al.(1995)]{Reid:1995} Reid, I.~N., Hawley, S.~L., \& Gizis, J.~E.\ 1995, \aj, 110, 1838 
\bibitem[Reid et al.(2007)]{Reid:2007} Reid, I.~N., Cruz, K.~L., \& Allen, P.~R.\ 2007, \aj, 133, 2825 
\bibitem[Rhee et al.(2007)]{Rhee:2007} Rhee, J.~H., Song, I., Zuckerman, B., \& McElwain, M.\ 2007, \apj, 660, 1556 
\bibitem[Riaz et al.(2006)]{Riaz:2006} Riaz, B., Gizis, J.~E., \& Harvin, J.\ 2006, \aj, 132, 866 
\bibitem[Rice et al.(2010)]{Rice:2010a} Rice, E.~L., Faherty, J.~K., \& Cruz, K.~L.\ 2010, \apjl, 715, L165 
\bibitem[Riedel et al.(2011)]{Riedel:2011} Riedel, A.~R., Murphy, S.~J., Henry, T.~J., et al.\ 2011, \aj, 142, 104 
\bibitem[Rodriguez et al.(2010)]{Rodriguez:2010} Rodriguez, D.~R., Kastner, J.~H., Wilner, D., \& Qi, C.\ 2010, \apj, 720, 1684
\bibitem[Rodriguez et al.(2011)]{Rodriguez:2011} Rodriguez, D.~R., Bessell, M.~S., Zuckerman, B., \& Kastner, J.~H.\ 2011, \apj, 727, 62 
\bibitem[Roeser et al.(2010)]{Roeser:2010} Roeser, S., Demleitner, M., \& Schilbach, E.\ 2010, \aj, 139, 2440 
\bibitem[Schlieder et al.(2012a)]{Schlieder:2012a} Schlieder, J.~E., L{\'e}pine, S., \& Simon, M.\ 2012a, \aj, 143, 80 % beta Pic & AB Dor members
\bibitem[Schlieder et al.(2012b)]{Schlieder:2012} Schlieder, J.~E., L{\'e}pine, S., Rice, E., et al.\ 2012b, \aj, 143, 114 % Na EW stuff
\bibitem[Schmitt et al.(1995)]{Schmitt:1995} Schmitt, J.~H.~M.~M., Fleming, T.~A., \& Giampapa, M.~S.\ 1995, \apj, 450, 392 %RASS limit cited here
\bibitem[Schneider et al.(2012a)]{Schneider:2012a} Schneider, A., Melis, C., \& Song, I.\ 2012a, \apj, 
754, 39
\bibitem[Schneider et al.(2012b)]{Schneider:2012b} Schneider, A., Song, I., Melis, C., Zuckerman, B., \& 
Bessell, M.\ 2012b, \apj, 757, 163 
\bibitem[Shkolnik et al.(2011)]{Shkolnik:2011} Shkolnik, E.~L., Liu, M.~C., Reid, I.~N., Dupuy, T., \& Weinberger, A.~J.\ 2011, \apj, 727, 6 
\bibitem[Shkolnik et al.(2012)]{Shkolnik:2012} Shkolnik, E.~L., Anglada-Escud{\'e}, G., Liu, M.~C., et al.\ 2012, \apj, 758, 56
\bibitem[Siebert et al.(2011)]{Siebert:2011} Siebert, A., Williams, M.~E.~K., Siviero, A., et al.\ 2011, \aj, 141, 187
\bibitem[Siess et al.(2000)]{Siess:2000} Siess, L., Dufour, E., \& Forestini, M.\ 2000, \aap, 358, 593 
\bibitem[Song et al.(2002)]{Song:2002} Song, I., Bessell, M.~S., \& Zuckerman, B.\ 2002, \apjl, 581, L43
\bibitem[Stauffer et al.(2007)]{Stauffer:2007} Stauffer, J.~R., Hartmann, L.~W., Fazio, G.~G., et al.\ 2007, \apjs, 172, 663 
\bibitem[Stauffer et al.(2010)]{Stauffer:2010} Stauffer, J., Tanner, A.~M., Bryden, G., et al.\ 2010, \pasp, 122, 885 
\bibitem[Stelzer et al.(2013)]{Stelzer:2013} Stelzer, B., Marino, A., Micela, G., Lopez-Santiago, J., \& Liefke, C.\ 2013, arXiv:1302.1061
\bibitem[Torres et al.(2000)]{Torres:2000} Torres, C.~A.~O., da Silva, L., Quast, G.~R., de la Reza, R., \& Jilinski, E.\ 2000, \aj, 120, 1410 
\bibitem[Torres et al.(2006)]{Torres:2006} Torres, C.~A.~O., Quast, G.~R., da Silva, L., de La Reza, R., Melo, C.~H.~F., \& Sterzik, M.\ 2006, \aap, 460, 695
\bibitem[Torres et al.(2008)]{Torres:2008} Torres, C.~A.~O., Quast, G.~R., Melo, C.~H.~F., \& Sterzik, M.~F.\ 2008, Handbook of Star Forming Regions, Volume II, 757 
\bibitem[Vacca et al.(2003)]{Vacca:2003} Vacca, W.~D., Cushing, M.~C., \& Rayner, J.~T.\ 2003, \pasp, 115, 389 % SpeXtool
\bibitem[Walkowicz et al.(2004)]{Walkowicz:2004} Walkowicz, L.~M., Hawley, S.~L., \& West, A.~A.\ 2004, \pasp, 116, 1105 %chi factor for L_Halpha
\bibitem[West et al.(2004)]{West:2004} West, A.~A., Hawley, S.~L., Walkowicz, L.~M., et al.\ 2004, \aj, 128, 426 
\bibitem[West et al.(2008)]{West:2008} West, A.~A., Hawley, S.~L., Bochanski, J.~J., et al.\ 2008, \aj, 135, 785 
\bibitem[White \& Basri(2003)]{White:2003} White, R.~J., \& Basri, G.\ 2003, \apj, 582, 1109
\bibitem[Wright et al.(2010)]{Wright:2010} Wright, E.~L., Eisenhardt, P.~R.~M., Mainzer, A.~K., et al.\ 2010, \aj, 140, 1868 
\bibitem[Yee \& Jensen(2010)]{Yee:2010} Yee, J.~C., \& Jensen, E.~L.~N.\ 2010, \apj, 711, 303 
\bibitem[Zacharias et al.(2004)]{Zacharias:2004} Zacharias, N., Monet, D.~G., Levine, S.~E., et al.\ 2004, Bulletin of the American Astronomical Society, 36, 1418 %NOMAD
\bibitem[Zacharias et al.(2010)]{Zacharias:2010} Zacharias, N., et al.\ 2010, \aj, 139, 2184 
\bibitem[Zacharias et al.(2012)]{Zacharias:2012} Zacharias, N., Finch, C.~T., Girard, T.~M., et al.\ 2012, VizieR Online Data Catalog, 1322, 0 
\bibitem[Zuckerman \& Webb(2000)]{Zuckerman:2000} Zuckerman, B., \& Webb, R.~A.\ 2000, \apj, 535, 959 
\bibitem[Zuckerman et al.(2001)]{Zuckerman:2001} Zuckerman, B., Song, I., \& Webb, R.~A.\ 2001, \apj, 559, 388 
\bibitem[Zuckerman \& Song(2004)]{ZS04} Zuckerman, B., \& Song, I.\ 2004, \araa, 42, 685
\bibitem[Zuckerman et al.(2006)]{Zuckerman:2006} Zuckerman, B., Bessell, M.~S., Song, I., \& Kim, S.\ 2006, \apjl, 649, L115 
\bibitem[Zuckerman et al.(2011)]{Zuckerman:2011} Zuckerman, B., Rhee, J.~H., Song, I., \& Bessell, M.~S.\ 2011, \apj, 732, 61 
\bibitem[Zuckerman \& Song(2012)]{Zuckerman:2012} Zuckerman, B., \& Song, I.\ 2012, \apj, 758, 77 

\end{thebibliography}
\end{document}